%% file: main.tex
\documentclass[acmtog]{acmart} 
\acmSubmissionID{267}

\usepackage{booktabs}

\setcitestyle{nosort,square} % nosort to allow for manual chronological ordering



\input{macros}

\acmJournal{TOG}
\acmYear{2020}\acmVolume{39}\acmNumber{4}\acmArticle{126}\acmMonth{7} \acmDOI{10.1145/3386569.3392415}
\begin{document}

\title{Point2Mesh: A Self-Prior for Deformable Meshes}

\author{Rana Hanocka}
\affiliation{\institution{Tel Aviv University}}

\author{Gal Metzer}
\affiliation{\institution{Tel Aviv University}}

\author{Raja Giryes}
\affiliation{\institution{Tel Aviv University}}

\author{Daniel Cohen-Or}
\affiliation{\institution{Tel Aviv University}}

\begin{abstract}
\input{abstract}
\end{abstract}

\input{figures/01_intro/teaser.tex}

\maketitle

\input{intro}
\input{relatedworks}
\input{overview}
\input{method}
\input{experiments}
\input{conclusion}

\bibliographystyle{ACM-Reference-Format}
\bibliography{bibs}

\end{document}

%% file: macros.tex
\usepackage{hyperref}
\usepackage{amsmath}
\usepackage{graphicx}
\usepackage{comment}
\usepackage{kbordermatrix}
\usepackage{multirow,bigdelim}

\usepackage{array}
\usepackage{wrapfig}

\usepackage{csvsimple}
\usepackage{adjustbox}
\usepackage[percent]{overpic}

\newcolumntype{C}[1]{>{\centering\let\newline\\\arraybackslash\hspace{0pt}}m{#1}}

\newcommand{\Mi}{\hat{M}_{l-1}}
\newcommand{\Mo}{\hat{M}_{l}}

\newcommand{\dVo}{\Delta \hat{V}_{l}}
\newcommand{\Vi}{\hat{V}_{l-1}}
\newcommand{\Vo}{\hat{V}_{l}}
\newcommand{\dEo}{\Delta \hat{E}_{l}}

\newcommand{\Wo}{W_{l}}
\newcommand{\partmesh}{PartMesh}

\newcommand{\ourmethod}{Point2Mesh}

\usepackage{pgfplots}
\pgfplotsset{compat=newest}
\usepgfplotslibrary{groupplots}
\usepgfplotslibrary{dateplot}
\usepackage[bottom]{footmisc}

%% file: abstract.tex
In this paper, we introduce \textit{\ourmethod{}}, a technique for reconstructing a surface mesh from an input point cloud. 
Instead of explicitly specifying a prior that encodes the expected shape properties, 
the prior is defined automatically using the input point cloud, which we refer to as a \emph{self-prior}. The self-prior encapsulates reoccurring geometric repetitions from a single shape within the weights of a deep neural network.
We optimize the network weights to deform an initial mesh to \emph{shrink-wrap} a single input point cloud. This explicitly considers the entire reconstructed shape, since shared local kernels are calculated to fit the overall object. The convolutional kernels are optimized globally across the entire shape, which inherently encourages local-scale geometric self-similarity across the shape surface. We show that shrink-wrapping a point cloud with a self-prior converges to a desirable solution; compared to a prescribed smoothness prior, which often becomes trapped in undesirable local minima. While the performance of traditional reconstruction approaches degrades in non-ideal conditions that are often present in real world scanning, \emph{i.e.,} unoriented normals, noise and missing (low density) parts, \ourmethod{} is robust to non-ideal conditions. We demonstrate the performance of \ourmethod{} on a large variety of shapes with varying complexity.

%% file: figures/01_intro/teaser.tex
\begin{teaserfigure}
\newcommand{\vfig}{18.5}
\centering
\includegraphics[width=\vfig cm]{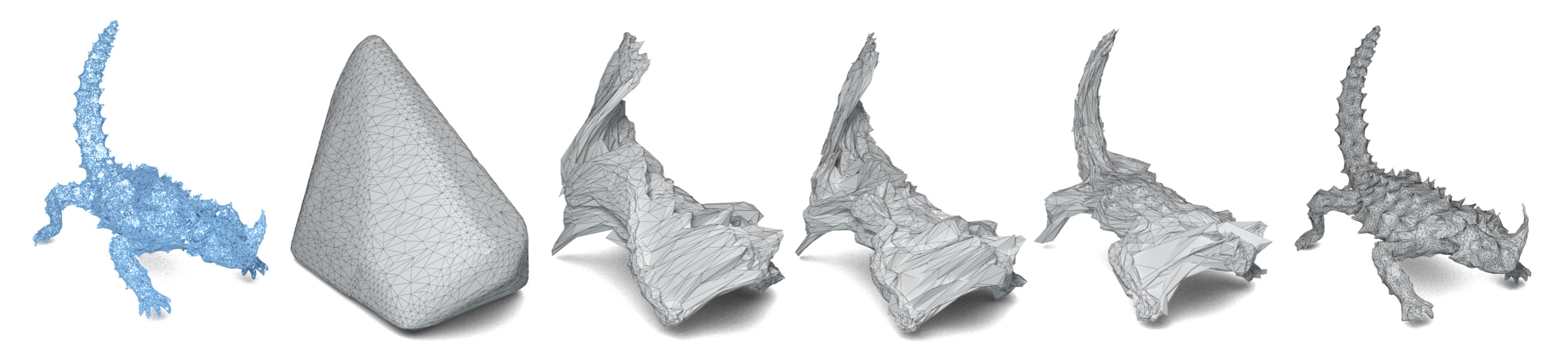}
\caption{
Starting with an input point cloud (left) and a deformable mesh, we iteratively shrink-wrap the input, leading to a watertight reconstruction.}
\label{fig:teaser}
\end{teaserfigure}

%% file: intro.tex
\section{Introduction}
Reconstructing a mesh from a point cloud is a long-standing problem in computer graphics. In recent decades, various approaches have been developed to reconstruct shapes for an array of applications~\cite{berger2017survey}.
The reconstruction problem is ill-posed, making it necessary to define a prior which incorporates the expected properties of the reconstructed mesh. Traditionally, priors are manually designed to encourage general properties, like piece-wise smoothness or local uniformity.

The recent emergence of deep neural networks carries new promise to bypass manually specified priors. Studies have shown that filtering data with a collection of convolution and pooling layers results in a salient feature representation, even with randomly initialized weights~\cite{saxe2011random,gaier2019weight}.
Convolutions exploit local spatial correlations, which are shared and aggregated across the entire data; while pooling reduces the dimensionality of the learned representation. Since shapes, like natural images, are not random, they have a distinct distribution which fosters the powerful intrinsic properties of CNNs to garner self-similarities.
 
In this paper, we introduce \textit{\ourmethod{}}, a method for reconstructing meshes from point clouds, where the prior is defined automatically by a convolutional neural network (CNN). 
Instead of explicitly specifying a prior, it is \textit{learned} automatically from a single input point cloud, without relying on any training data or pre-training, in other words, a \emph{self-prior}. 
In contrast, supervised learning paradigms demand large amounts of input (point cloud) and ground-truth (surface) training pairs (\emph{i.e., data-driven priors}), which often entails modeling the acquisition process.
An appealing aspect of \ourmethod{} is that it does not require knowing the distribution of noise or partiality during training, bypassing the need for large amounts of labeled training data from a particular distribution. Instead, \ourmethod{} optimizes a CNN-based self-prior during inference time, which leverages the self-similarity present within a single shape.
\input{figures/06_experiments/068_self_similarity/figure.tex}

To reconstruct a mesh, we take an optimization approach, where an initial mesh is iteratively deformed to fit the input point cloud by a series of steps conducted by a CNN (Figure~\ref{fig:teaser}). Since objects are solid, the reconstructed model must be watertight. Therefore, we start with a watertight mesh, and deform it by displacing the vertex positions in incremental steps, to preserve the required watertight property. The initial deformable mesh can be estimated using different techniques that roughly approximate the input shape with an arbitrary genus.

Neural networks have an implicit tendency to learn regularized weights~\cite{achille2018emergence}, and converge to low-entropy solutions even for single images~\cite{DoubleDIP}.
Since \ourmethod{} encodes the entire shape into the parameters of a CNN, it leverages the representational power of networks to inherently remove noise and outliers. Accordingly, 
this powerful self-prior excels at modeling natural shapes, whereas noisy and unnatural shapes are not well explained by the CNN (see Figure~\ref{fig:network_prior}). A notable advantage of \ourmethod{}, especially when compared to classic techniques, is its ability to cope with non-ideal conditions which are often present in real world scanning, \emph{i.e.,} unoriented normals, noise and / or missing regions. The classic Poisson surface reconstruction~\cite{kazhdan2006poisson} technique assumes oriented normals, a condition which is often hard to meet \cite{huang2009consolidation}.

\ourmethod{} optimization relies on MeshCNN~\cite{Hanocka2019MeshCNN}, which offers the platform for a convolutional neural network applied on meshes. 
MeshCNN learns convolutional kernels directly on the edges of the mesh for tasks like classification and segmentation. In this work, we extend the capabilities of MeshCNN to handle shape regression.
To best \emph{explain} the input shape, the CNN weights must leverage the self-similarity present in natural shapes, by aggregating local attributes specific to the entire reconstructed shape. On the one hand, convolutions are applied locally to extract salient features; on the other, the same local kernels are utilized over the whole shape. Optimizing kernel weights globally across the entire shape, inherently encourages local-scale geometric self-repetition across the shape surface. Accordingly, the \textit{structure} of a CNN inherently encapsulates the essence of natural shapes, which we leverage as a self-prior for reconstructing mesh surfaces. 

We demonstrate the performance of \ourmethod{} on variety of shapes with varying complexity, including real scans obtained from a NextEngine 3D laser scanner.
We demonstrate the applicability of our approach to reconstruct a mesh from point samples containing noise, unoriented / unknown normals and / or missing regions, and compare to several other leading techniques. In particular, we demonstrate the advantage of a network over the ubiquitous smoothness prior, and pure optimization (of the same objective) with no network prior, to emphasize the strength of our self-prior. 

%% file: figures/06_experiments/068_self_similarity/figure.tex
\begin{figure}[h]
    \centering
    \newcommand{\pl}{-2}
    \begin{overpic}[width=8.5cm]{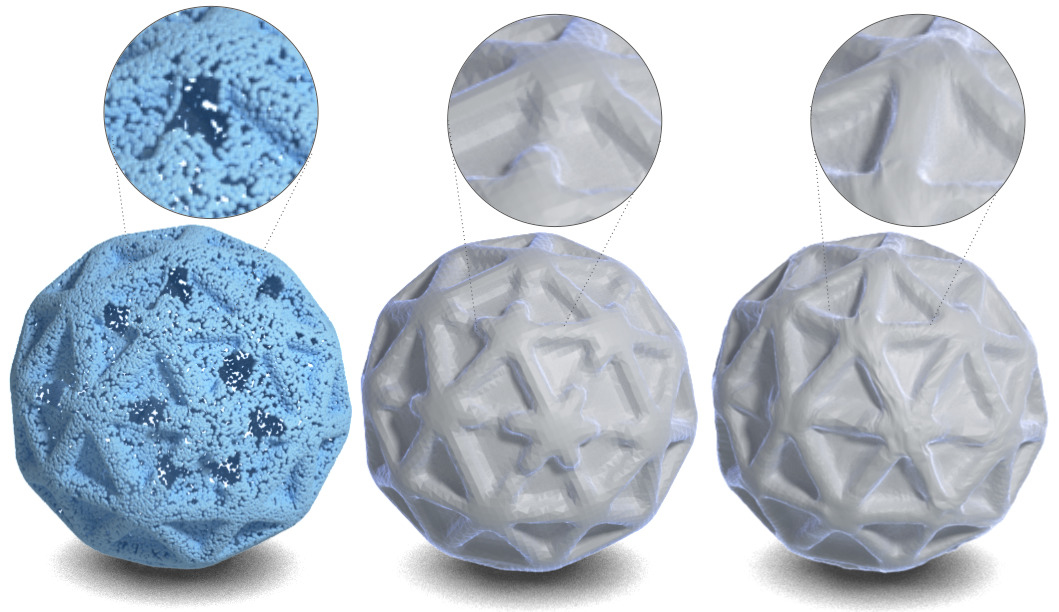}
    \put(12,  \pl){\textcolor{black}{Input}}
    \put(40.5, \pl){\textcolor{black}{Smooth-prior}}
    \put(76, \pl){\textcolor{black}{Self-prior}}
    \end{overpic}
    \vspace{-5pt}
    \caption{Reconstructing a complete mesh from a point cloud with missing regions using a smooth-prior ignores the character of the global shape. The self-prior in \ourmethod{} inherently learns and leverages the recurrences present within a single shape, leading to a more plausible reconstruction.}
    \label{fig:selfsim}
\end{figure}

%% file: relatedworks.tex
\input{figures/03_overview/031_noise_prior_landscape.tex}
\section{Related Work}
% \textbf{Surface Reconstruction.} 
\subsubsection*{Surface Reconstruction}
There has been a lot of research on the inverse problem of reconstructing a surface from a point cloud. Early works approached the reconstruction problem by fitting implicit functions to the point cloud \cite{Hoppe:1992}, and then generating a mesh over its zero-set. Most previous works on surface reconstruction have focused on data consolidation 
\cite{ohtake2003multi, lipman2007parameterization, huang2009consolidation, miao2009shape, Oztireli10Spectral, huang2013edge}, 
that is, denosing, smoothing, up-sampling, or generally, enhancing the given point cloud. 

The most popular method for surface reconstruction is Poisson reconstruction~\cite{kazhdan2006poisson, kazhdan2013screened}, which formalizes the problem of surface reconstruction as a Poisson system. Finding an indicator function whose gradient is the vector field characterized by the sample points and normals, enables reconstruction of the surface. Alternative approaches define the indicator function using Fourier coefficients~\cite{kazhdan2005reconstruction} and Wavelets~\cite{manson2008streaming}.
Another class of approaches partitions the implicit function into multiple scales~\cite{ohtake2003multi,nagai2009smoothing}. Ohtake et al.~\shortcite{ohtake20053d} use the compact radial basis functions (RBFs) to build the implicit surface.

Poisson reconstruction, like most previous approaches, processes point sets in local regions, in order to facilitate a post-process triangulation and mesh reconstruction. These methods do not enforce global constraints on the generated mesh, and, in particular, they require perfect normal orientation (\textit{i.e.,} each normal points outside of the surface) to guarantee that the reconstructed surface is watertight. Since perfect normal orientation is a cumbersome requirement, an alternative approach~\cite{hornung2006robust} relaxes this requirement by computing a dilated voxel crust and extracts the final closed surface via graph-cut minimization.

Surface reconstruction has also been approached via mesh deformation, whereby an initial mesh is iteratively deformed to fit a given point cloud~\cite{sharf2006competing, li2010analysis}.
This approach is analogous to \textit{active contours}, also referred to as Snakes~\cite{kass1988snakes, xu1998snakes, mcinerney2000t}. Such optimizations are highly non-convex and often lead to local minima. To alleviate the inherent ambiguities, such techniques must define strong priors, which are explicitly hand-crafted, for example, encouraging locally uniform and piece-wise smooth reconstructions.

The most related work to \ourmethod{} is the work of Sharf et al. ~\shortcite{sharf2006competing}. Their approach reconstructs a watertight surface, via a series of iterative optimizations. Starting from an initial spherical mesh placed within the point cloud, the mesh is deformed in the direction of the face normals to fit the target point cloud. Unlike ~\cite{sharf2006competing}, where the mesh is inflated starting inside and moving out, we approach the mesh fitting from outside-in. To avoid local minima, Sharf et al. used smoothness priors and heuristics to split and prioritize the advancing fronts. The key meta difference of our method (beyond technical details) is that we do not design the prior; instead, we \textit{learn} the prior from the input shape itself. 

\subsubsection*{Neural Self-Priors}
% \textbf{Deep Neural Networks for Self-Priors.}
The inspiration for \ourmethod{} comes from 2D image processing, in particular the work of Ulyanov et al. ~\shortcite{ulyanov2018deep} and other works that focus on self-similarity within images, \emph{e.g.,}~\cite{shocher2018zero, DoubleDIP, Zhou2018, shaham2019singan, sun2019test}. The concept of Deep Image Prior (DIP)~\cite{ulyanov2018deep} is intriguing, yet, still not fully understood. In our work, we not only use a similar concept for surface reconstruction, but show empirically that the weight-sharing of the CNN network is a key element in the advantage of self priors. We believe that this directs the network to learn (non-local) repeating structures within the processed data (mesh, in our case). The notion of exploiting symmetry and repetitions in 3D shapes has been explored in classic shape analysis techniques~\cite{nan2010smartboxes, pauly2008discovering}. \citet{harary2014context} demonstrated surface holes can be completed using patches from within the same object.

% \textbf{Geometric Deep Learning and Reconstruction.}
\subsubsection*{Geometric Deep Learning and Reconstruction}
Recently, the concept of a differential 3D rendering as a \textit{plug-and-play} component in a computer graphics pipeline has gained increasing interest. An application of a differential rendering module is reconstructing a 3D model, textures and lighting from the rendered image. In DIB-R~\cite{chen2019dibrender} an interpolation-based differential renderer is proposed, to reconstruct color and geometry from an RGB image. Yifan el al. \shortcite{Yifan:DSS:2019} propose a surface splatting technique, using a a point-based differential renderer. Recently, several data-driven techniques for reconstructing a 3D mesh from an image using neural networks have been proposed~\cite{wang2018pixel2mesh, kurenkov2018deformnet, wen2019pixel2mesh++}. Scan2Mesh~\cite{dai2019scan2mesh} proposed a data-driven technique for reconstructing a complete mesh from an input scan, and represent the partial input mesh as a TSDF. SDM-Net~\cite{gao2019sdm} trained a neural network to deform mesh cuboids for shape synthesis. For a survey on geometric deep learning the reader is referred to~\cite{bronstein2017geometric,xiao2020survey}.

The recent work of Williams et al.~\shortcite{williams2019deep} presents Deep Geometric Prior (DGP). Despite the apparent similarity, it shares little commonality with our work. Similar to DIP and \ourmethod{}, they train a network (based on AtlasNet~\cite{groueix2018papier}) from a single input point cloud. However, they train different MLPs for each local region in the point cloud, which is used to reconstruct local (disjoint) charts. These local charts are used to upsample and enhance the sampled points. The final mesh is reconstructed in a post-process by sampling the local charts and then using Poisson reconstruction. It should be stressed that unlike \ourmethod{}, Deep Geometric Prior does not use weight sharing among the separate MLPs, and thus lacks global self-similarity analysis. In fact, DGP is closer in spirit to the recent works that use neural networks for point up-sampling (\emph{e.g.,}~\cite{yu2018pu, li2019pugan}). \ourmethod{} builds upon recent advances in 3D mesh deep learning.  In particular, we use MeshCNN~\cite{Hanocka2019MeshCNN}, which is an edge-based CNN with weight-sharing convolutions, and learnable pooling capabilities, enabling learning a self-prior for surface reconstruction.

%% file: figures/03_overview/031_noise_prior_landscape.tex
\begin{figure*}
\begin{tabular}{cc}
\hspace{8pt}
\resizebox{5.2cm}{!}{\input{figures/03_overview/031_noise_prior_graph.tex}} &
\begin{overpic}[width=10cm]{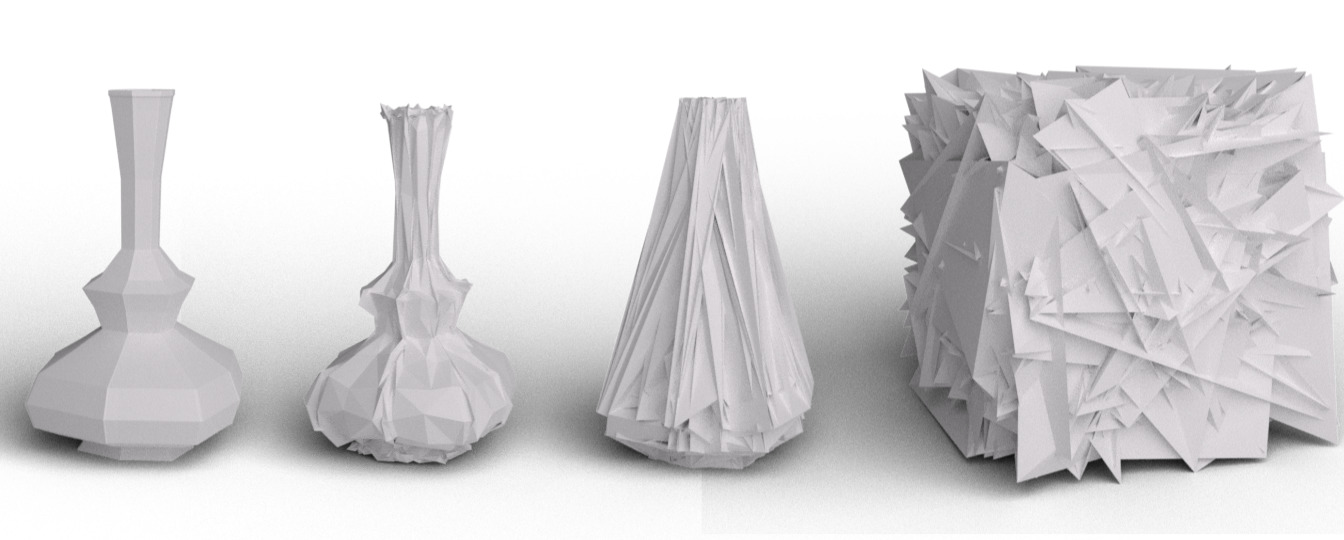}
    \put(8.5,2.5){\textcolor{black}{(a)}}
    \put(30,2.5){\textcolor{black}{(b)}}
    \put(52,2.5){\textcolor{black}{(c)}}
    \put(85,2.5){\textcolor{black}{(d)}}
\end{overpic} \\
\end{tabular}
\vspace{-10pt}
\caption{The CNN structure inherently prefers reconstructing \textit{natural} shapes. Graph of reconstruction error vs. optimization iterations (left) for reconstructing four different shapes: (a) natural shape, (b) shape + noise, (c) shuffled vertices and (d) uniform noise. The network is able to best reconstruct the natural shape (a), and struggles to reconstruct noisy and chaotic (\emph{unnatural}) shapes (b,c,d).}
\label{fig:network_prior}
\end{figure*}
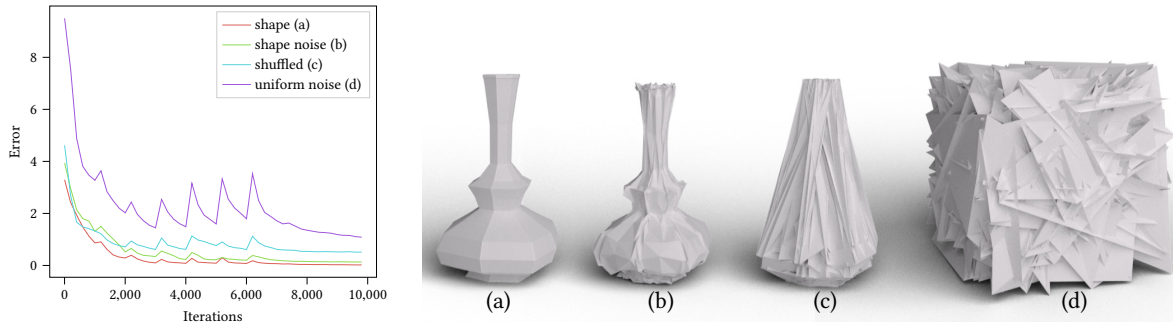

%% file: figures/03_overview/031_noise_prior_graph.tex
% This file was created by tikzplotlib v0.8.7.
\begin{tikzpicture}

\definecolor{color0}{rgb}{0.86,0.3712,0.34}
\definecolor{color1}{rgb}{0.5688,0.86,0.34}
\definecolor{color2}{rgb}{0.34,0.8288,0.86}
\definecolor{color3}{rgb}{0.6312,0.34,0.86}

\begin{axis}[
legend cell align={left},
legend style={fill opacity=0.8, draw opacity=1, text opacity=1, draw=white!80.0!black},
scaled x ticks=false,
tick align=outside,
tick pos=left,
x grid style={white!69.01960784313725!black},
xlabel={Iterations},
xmin=-489, xmax=10291,
xtick style={color=black},
y grid style={white!69.01960784313725!black},
ylabel={Error},
ymin=-0.454334897175431, ymax=9.97237045057118,
ytick style={color=black}
]
\addplot [semithick, color0]
table {%
1 3.29232358932495
201 2.414142370224
401 1.93066155910492
601 1.47830975055695
801 1.13094294071198
1001 0.867908596992493
1201 0.905170679092407
1401 0.618748605251312
1601 0.405022591352463
1801 0.32147204875946
2001 0.28699791431427
2201 0.390965431928635
2401 0.252503097057342
2601 0.170215994119644
2801 0.122237108647823
3001 0.0999931767582893
3201 0.232444763183594
3401 0.137319713830948
3601 0.112210966646671
3801 0.0997837260365486
4001 0.0812297239899635
4201 0.271574974060059
4401 0.127613246440887
4601 0.110764883458614
4801 0.0986851155757904
5001 0.0861332416534424
5201 0.298598766326904
5401 0.129264160990715
5601 0.0973243489861488
5801 0.0870006680488586
6001 0.0768511891365051
6201 0.179632052779198
6401 0.109978683292866
6601 0.0844224244356155
6801 0.0729832723736763
7001 0.0693109109997749
7201 0.0574575215578079
7401 0.0602700039744377
7601 0.0468044430017471
7801 0.0388026610016823
8001 0.0391211584210396
8201 0.0344365686178207
8401 0.0340683460235596
8601 0.0339826419949532
8801 0.0292522758245468
9001 0.0290679633617401
9201 0.0275907441973686
9401 0.0241023004055023
9601 0.0207308977842331
9801 0.0196062549948692
};
\addlegendentry{shape (a)}
\addplot [semithick, color1]
table {%
1 3.93902277946472
201 2.95743465423584
401 2.13608956336975
601 1.80084276199341
801 1.7061995267868
1001 1.31382656097412
1201 1.5049022436142
1401 1.24194145202637
1601 1.01688396930695
1801 0.789372503757477
2001 0.52185583114624
2201 0.654087722301483
2401 0.473060935735703
2601 0.389266908168793
2801 0.367635786533356
3001 0.343167454004288
3201 0.554921090602875
3401 0.463455677032471
3601 0.374138385057449
3801 0.261018693447113
4001 0.231462121009827
4201 0.496633619070053
4401 0.391921043395996
4601 0.251975774765015
4801 0.219670102000237
5001 0.223045468330383
5201 0.307276844978333
5401 0.246893748641014
5601 0.231946334242821
5801 0.211794093251228
6001 0.19980376958847
6201 0.383019655942917
6401 0.330510377883911
6601 0.268274247646332
6801 0.220945537090302
7001 0.200281053781509
7201 0.178391203284264
7401 0.170876458287239
7601 0.157826870679855
7801 0.159881636500359
8001 0.151075437664986
8201 0.146907404065132
8401 0.142396450042725
8601 0.141960993409157
8801 0.136752650141716
9001 0.14018751680851
9201 0.140161469578743
9401 0.134627491235733
9601 0.133334934711456
9801 0.131875678896904
};
\addlegendentry{shape noise (b)}
\addplot [semithick, color2]
table {%
1 4.61943626403809
201 2.68786835670471
401 1.65975618362427
601 1.48284959793091
801 1.41606426239014
1001 1.32569909095764
1201 1.2186621427536
1401 0.981513381004334
1601 0.840973019599915
1801 0.762320816516876
2001 0.716007888317108
2201 0.938131988048554
2401 0.792852282524109
2601 0.732177913188934
2801 0.663280487060547
3001 0.616226851940155
3201 1.05307674407959
3401 0.783037722110748
3601 0.717027008533478
3801 0.649641931056976
4001 0.619180142879486
4201 1.13167548179626
4401 0.979165315628052
4601 0.918988883495331
4801 0.835343480110168
5001 0.764806389808655
5201 0.903238952159882
5401 0.743618845939636
5601 0.685883641242981
5801 0.660188913345337
6001 0.610257208347321
6201 1.12029898166656
6401 0.876220226287842
6601 0.748075187206268
6801 0.689071774482727
7001 0.618938744068146
7201 0.597834348678589
7401 0.590205550193787
7601 0.58091002702713
7801 0.541249632835388
8001 0.541808128356934
8201 0.529928863048553
8401 0.526109874248505
8601 0.533226251602173
8801 0.526326477527618
9001 0.519267678260803
9201 0.519940674304962
9401 0.52779632806778
9601 0.51130199432373
9801 0.51181161403656
};
\addlegendentry{shuffled (c)}
\addplot [semithick, color3]
table {%
1 9.49842929840088
201 7.54173469543457
401 4.87092542648315
601 3.81114840507507
801 3.47102427482605
1001 3.27356290817261
1201 3.63905453681946
1401 2.83296656608582
1601 2.49047517776489
1801 2.200279712677
2001 2.02435970306397
2201 2.43574857711792
2401 1.96389186382294
2601 1.72078311443329
2801 1.54279220104218
3001 1.44404256343842
3201 2.53953814506531
3401 2.05803442001343
3601 1.77607119083405
3801 1.60142397880554
4001 1.49037837982178
4201 3.15411257743835
4401 2.33847665786743
4601 1.93711841106415
4801 1.76644897460938
5001 1.5953311920166
5201 3.32440090179443
5401 2.54645824432373
5601 2.21874380111694
5801 2.01240038871765
6001 1.79446136951447
6201 3.52216219902039
6401 2.49528098106384
6601 2.0361065864563
6801 1.88407158851624
7001 1.71672403812408
7201 1.59934937953949
7401 1.62879478931427
7601 1.51753318309784
7801 1.40423965454102
8001 1.35423719882965
8201 1.31204319000244
8401 1.27416932582855
8601 1.26252007484436
8801 1.23708081245422
9001 1.18597972393036
9201 1.15695703029633
9401 1.15355050563812
9601 1.10853612422943
9801 1.08398604393005
};
\addlegendentry{uniform noise (d)}
\end{axis}

\end{tikzpicture}

%% file: overview.tex
\input{figures/03_overview/030_arch.tex}
\vspace{-10pt}
\section{Point2Mesh}
\ourmethod{} reconstructs a watertight mesh by optimizing the weights of a CNN to iteratively deform an initial mesh to shrink-wrap the given input point cloud $X$ (examples in Figure~\ref{fig:convergence}). The CNN automatically defines the self-prior, which enjoys the innate properties of the network structure. Key to the self-prior is the weight sharing structure of the CNN, which inherently models recurring and correlated structures and, hence, is weak in modeling noise and outliers, which have non-recurring geometries. Natural shapes contain strong self-correlation across multiple-scales and fine-grained details often reoccur, while noise is random and uncorrelated; making reconstructing recurring fine-grained details while eliminating noise possible. Observe Figure~\ref{fig:anky}, where the self-prior removes the
noisy bumps on the reoccurring ridges on the Ankylosaurus back and tail, yet still preserves the fine-grained ridges on the neck. Moreover, notice how the self-prior complete the missing portions in Figure~\ref{fig:selfsim} and~\ref{fig:flat_missing} in a manner that is consistent with the characteristic of the global shape. An overview of our approach is illustrated in Figure~\ref{fig:overview}.

We perform the reconstruction in a coarse-to-fine manner, where the output mesh at level $l$ is denoted by $\Mo{}$.
The network takes as input a mesh $\Mi{}$ and predicts displacements $\dVo{}$, which move the mesh vertices towards the point cloud, driven by a loss to close the gap between the mesh surface and the point cloud. 
The convergence is iterative, since the displacements continue to increase to deform the mesh to shrink wrap the point cloud.
In each iteration the network receives as input a random noise vector $C_l$, and the weights $\Wo{}$ are randomly initialized.  
The $C_l$ vector serves as 
a random initialization for the predicted displacements, and the network weights are optimized to minimize the loss, while $C_l$ remains constant. 

\ourmethod{} works in coarse-to-fine \textit{levels}, where in each level the predicted mesh is refined by subdividing its triangles. The initial mesh is coarse approximation of the point cloud. If the object has a genus of zero we use the convex hull of the point cloud, whereas we use a coarse alpha shape~\cite{edelsbrunner1994three} for shapes with arbitrary genus (alternatively, we can use other methods).

Our network is comprised of a series of \textit{edge-based} convolution and pooling layers, as originally proposed in MeshCNN. However, unlike MeshCNN, which uses geometric input features, \ourmethod{} input features are the entries of the random (fixed) vector $C_l$. The convolutional kernels operate on the edges of the input mesh $\Mi{}$, initialized by $C_l$, and are optimized to predict vertex displacements $\dVo{}$. Specifically, the edge filters regress a list of displaced mesh edges $\dEo{}$, where each edge is represented by the three coordinates of both endpoint vertices (dimension $\#E \times 2 \times 3$). The location of each vertex in the list of edges is averaged to yield the final list of per-vertex displacements $\dVo{}$.
\input{figures/04_method/040_build_v.tex}

Once the displacements $\dVo{}$ are predicted, the updated mesh $\Mo{}$ is reconstructed, and the loss is computed by measuring the gap between the updated mesh and the point cloud, which drives the optimization of the network filters. Note that since the vertex displacements do not alter the mesh connectivity, the reconstructed mesh $\Mo{}$ has the prescribed connectivity and genus of $\Mi{}$. 

The input tensor $C_l$ has six feature vectors per edge, with the same number of edges as $\Mi{}$, sampled from a uniform distribution $[0, 1)$. The values of $C_l$ are sampled once per level, and remain constant during the optimization process. Effectively, the network acts as a parameterization of the input point cloud $X$ by the network weights and the fixed vector $C_l$, \emph{i.e.,} $X = f_{\Wo{}}(C_l)$. However, we often do not actually wish to recover $X$ exactly, since it can be corrupted with noise, unoriented normals or missing regions. 
Setting $C_l$ to be random helps the network avoid bias and overfitting.

%% file: figures/03_overview/030_arch.tex
\begin{figure*}[ht]
    \centering
    \newcommand{\pl}{6}
    \newcommand{\vl}{8}
    \begin{overpic}[width=18cm]{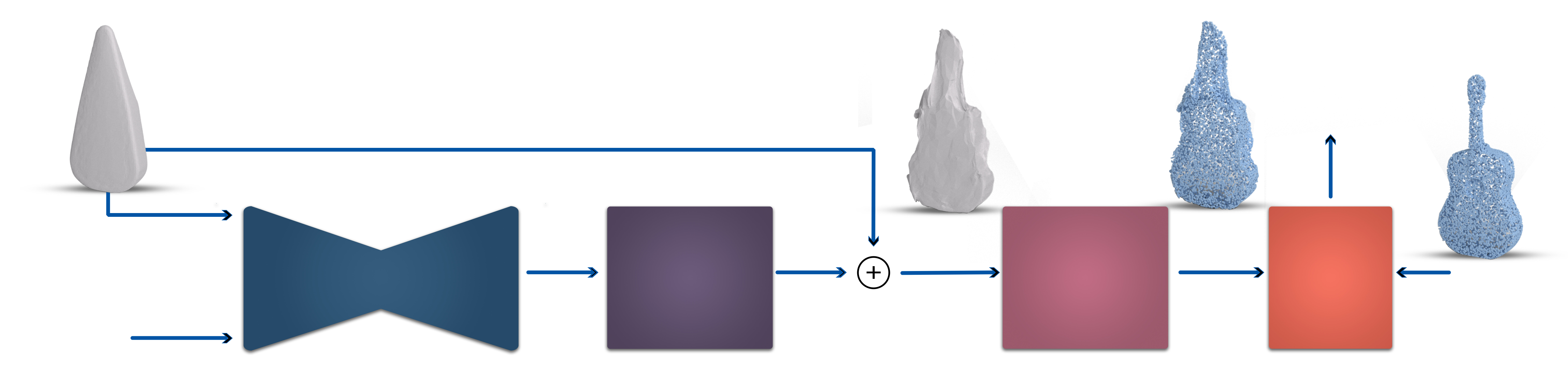}
    \put(5, 9.5){\textcolor{black}{$M_{l-1}$}}
    \put(5, 2.8){\textcolor{black}{$C_{l-1}$}}
    \put(21,\pl){\textcolor{white}{Self-Prior}}
    \put(40.8,\pl){\textcolor{white}{Build $\Delta V$}}
    \put(66.1,\pl){\textcolor{white}{Sampler}}
    \put(83.5,\pl){\textcolor{white}{Dist}}
    \put(34,\vl){\textcolor{black}{$\Delta \hat{E}_l$}}
    \put(50,\vl){\textcolor{black}{$\Delta \hat{V}_l$}}
    \put(60,\vl){\textcolor{black}{$\hat{M}_l$}}
    \put(77,\vl){\textcolor{black}{$\hat{Y}_l$}}
    \put(94, 6){\textcolor{black}{$X$}}
    \put(83, 16.5){\textcolor{black}{Loss}}
    \end{overpic} \\
    \caption{Overview of \ourmethod{} framework in a level $l$. The initial mesh $M_{l-1}$ and fixed random constant $C_{l-1}$ are input to the network (self-prior), which outputs a differential displacement vector per edge $\Delta \hat{E}_l$. The differential displacement per vertex $\Delta \hat{V}_l$ is calculated by averaging the displacements for each of its incident edges. The reconstructed mesh $\hat{M}_l$ has vertices given by the vertices of $\hat{M}_l$ plus $\Delta \hat{V}_l$, and the connectivity of $\hat{M}_l$. The reconstructed mesh is sampled to get $\hat{Y}_l$ which is compared against the input point cloud $X$. This loss is back-propagated in order to update self-prior network weights.}
    \label{fig:overview}
\end{figure*}
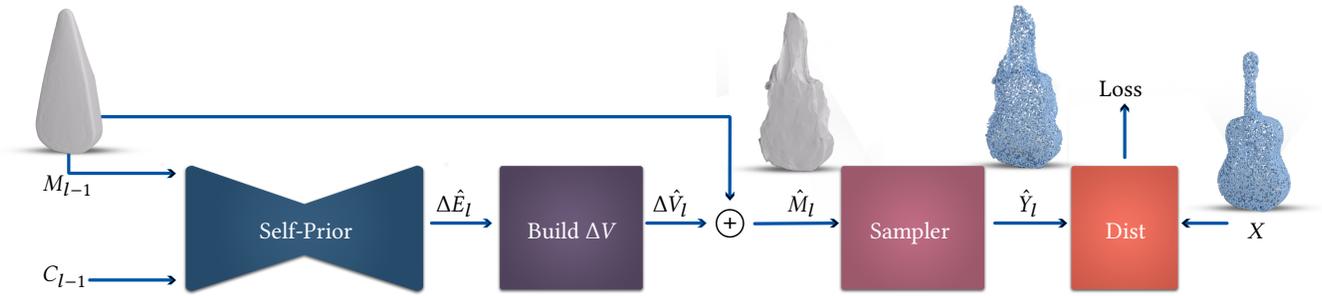

%% file: figures/04_method/040_build_v.tex
\begin{figure}[b]
    \centering
    \adjincludegraphics[width=5.5cm,trim={{0.11\width} {.13\height} {.02\width} {.17\height}},clip]{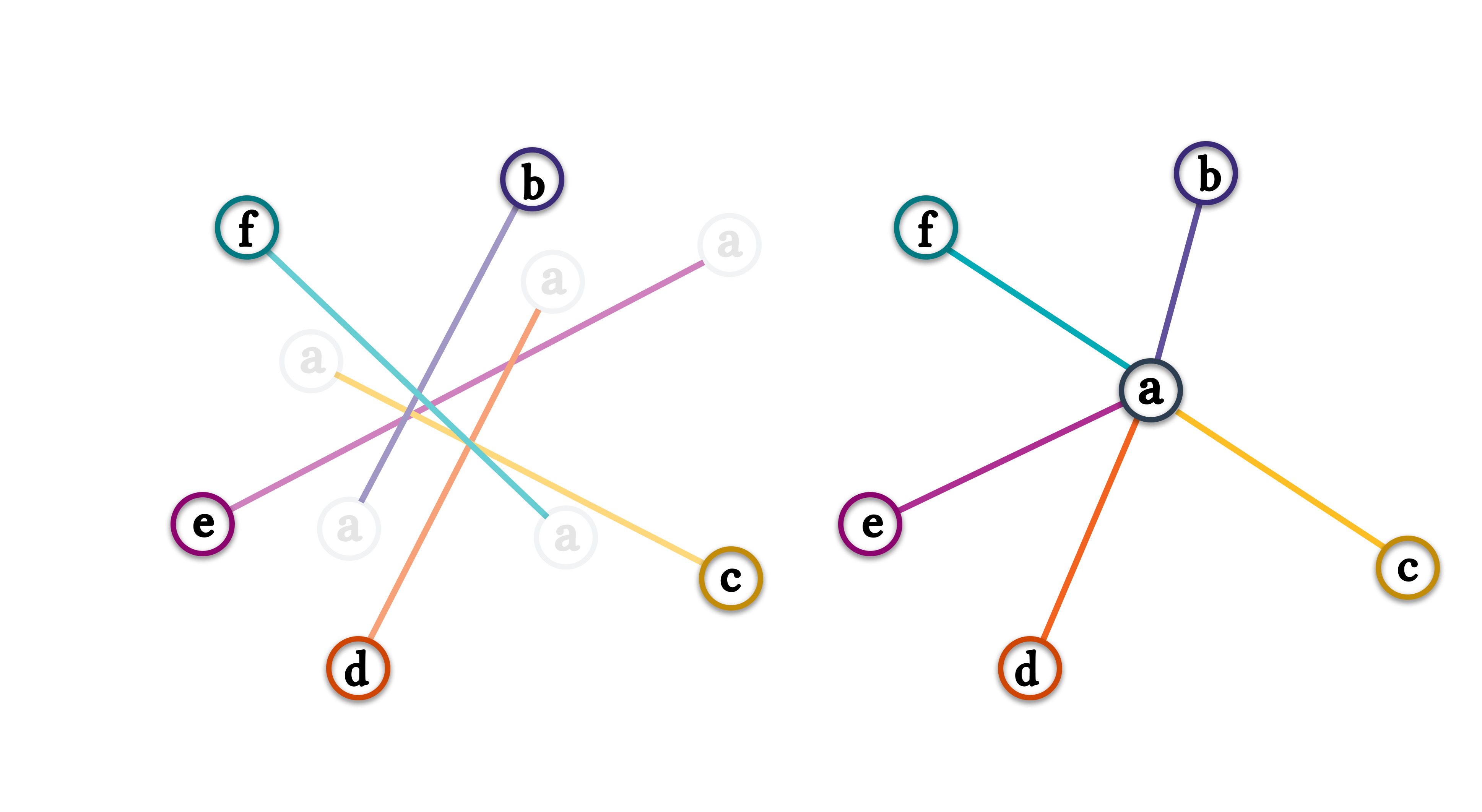}
    \caption{Illustration of edge displacements predicted by network (left) and then averaged (right).}
    \label{fig:build_v}
\end{figure}

%% file: method.tex
\input{figures/06_experiments/068_self_similarity/anky.tex}
\subsection{Explicit Mesh Representation}
The reconstructed surface is represented as a triangular mesh. A triangular mesh is defined by a set of vertices, (triangular) faces and edges: ($V$, $F$, $E$), where $V= \left \{ \mathbf{v}_1, \mathbf{v}_2 \cdots \right \}$ is the set of vertex positions in $\mathbb{R}^3$. The connectivity is given by a set of faces $F$ (triplets of vertices), and a set of edges $E =\left \{ \mathbf{e}_1, \mathbf{e}_2 \cdots \right \}$ (pairs of vertices).

The reconstructed surface $\Mo{}$ is directly deformed by our network. Instead of estimating the \emph{absolute} vertex position $\Vo{}$ of the deformed mesh, the network estimates a \emph{differential} vertex position $\dVo{}$, relative to the input mesh vertices $\Vi{} \in \Mi{}$ (\emph{i.e.,} $\Vo{} = \Vi{} + \dVo{}$). This facilitates learning in two main ways. First, the last layer in the network can be initialized with zeros 
(\emph{i.e.,} no displacement), which leads to a better initial condition for the optimization. Second, there is evidence in neural networks that predicting residuals yields more favorable results \cite{Shamir18ResNets}, especially for tasks which regress displacements~\cite{hanocka2018alignet}.

We use the neural building blocks of MeshCNN to regress the final vertex locations of the predicted mesh. 
However, since MeshCNN operates on \textit{edges}, it regresses a list of edge displacements $\dEo{}$, consisting of pairs of vertex displacements. Since a particular vertex is incident to many edges (\emph{i.e.,} six on average), the final vertex locations must aggregate over all the predicted vertex locations (illustrated in Figure~\ref{fig:build_v}). 
To handle this, the over-determined list of vertex displacements in $\dEo{}$ is fed into a \emph{Build $\Delta V$} module (Figure~\ref{fig:overview}) that creates a unique list of vertex displacements $\dVo{}$.
Specifically, for each vertex, the predicted vertex displacements in each edge are averaged to get the final list of vertices. 
A particular vertex position $\mathbf{\hat{v}}_i$ is calculated by
\begin{equation}
    \mathbf{\hat{v}}_i = \frac{1}{n} \sum_{j \in n} \mathbf{\hat{e}_j}(v_{i}),
\end{equation}
where $\mathbf{\hat{e}_j}(v_{i})$ is the vertex position of $v_{i}$ in edge ${e}_j$, and $n$ is the valence of vertex $\mathbf{\hat{v}}_i$.

\subsection{Mesh to Point Cloud Distance}
The estimated mesh vertices are driven by the distance of the deformed mesh $\Mo{}$ to the input point cloud $X$. The deformed mesh $\Mo{}$ is sampled via a (differential) sampling layer resulting in a point set $\hat{Y}$ that is measured against the input point set $X$. We evaluate the distance using the bi-directional Chamfer distance:
\begin{equation}
    d(X,\hat{Y}) = \sum_{x \in X} \min_{\hat{y} \in \hat{Y}} || x - \hat{y} ||_2 + 
    \sum_{\hat{y} \in \hat{Y}} \min_{x \in X} || x - \hat{y} ||_2,
\end{equation}
which is fast, differential and robust to outliers~\cite{fan2017point}.
%
% \vspace{-5pt}
% \textbf{Differential Sampler.} 
% \subsubsection{Differential Sampler}

To compare the reconstructed mesh to the input point cloud, we sample the mesh surface. Since the network weights encode information, which ultimately predicts the deformed mesh vertex locations, the sampling mechanism must be differential in order to back-propagate gradients through the network weights. 
A point sampled from a particular triangle has some distance to the closest point in the point cloud. This distance contributes to a gradient update, which will displace the three vertices that define the face of the sampled point. A mesh can be sampled by first selecting a face from a distribution $P_f$ and then sampling a point within each triangle from another distribution $P_p$. We use a uniform sampler, where the area of each triangle is proportional to the probability of selecting it $P_f \propto A_f$. Uniformly sampling a point inside a triangle with 
vertices $v_0, v_1, v_2$ (where $v_0$ is at the origin), is given by ${a_1 \cdot v_1 + a_2 \cdot v_2}$ (where $a_1$ and $a_2$ are $\in U(0,1)$ and $a_1 + a_2 < 1$).

\input{figures/06_experiments/068_self_similarity/flat_missing.tex}
\subsection{Beam-Gap Loss}
Deforming a mesh to enter narrow and deep cavities is a difficult task. Specifically, a mesh optimization which only uses the bi-directional Chamfer distance can become trapped in a local minimum, without ever entering the cavity. To drive the mesh into deep cavities, we calculate beam rays, from the mesh to the input point cloud, which we call \textit{beam-gap}. Calculating the exact beam-gap intersection is not possible, since the input point cloud is a sparse and discrete representation of a continuous surface. Therefore, we propose a method for approximating the distance to the underlying surface, which serves as an additional loss to the Chamfer loss.

For a point $\hat{y}$ sampled from a deformable mesh face, a beam is cast in the direction of the deformable face normal.
The beam intersection with the point cloud is the closest point in an $\epsilon$ cylinder around the beam.
More formally, the beam intersection (or collision) of a point $\hat{y}$ is given by $\mathcal{B}(\hat{y}) = x$, and so the overall beam-gap loss is given by:
\begin{equation}
    \mathcal{L}_b(X, \hat{Y}) = \sum_{\hat{y} \in \hat{Y}} \left\lVert {\hat{y} - \mathcal{B}(\hat{y})} \right\rVert ^2.
\end{equation}
Since the underlying surface of the target shape is unknown, the beam-gap distance is a discrete approximation of the true beam intersection. Our system is not sensitive to the selection of $\epsilon$ (used $\epsilon = 0.99$), assuming a reasonable sampling density.

\input{figures/06_experiments/062_converge/figure.tex}
Note, that not all points on the reconstructed mesh obtain a beam collision to the point cloud. In this case, the beam-gap of $\mathcal{B}(\hat{y}) = \hat{y}$ (\emph{i.e.,} no penalty). Furthermore, if multiple beam intersections are detected, we select the closest one. If for a given point sampled from the reconstructed mesh there is a \textit{good fit} to the target point cloud, then there is no beam-gap penalty for this point. We determine areas with a good fit by computing the mutual $k$-Nearest Neighbor ($k$-NN) between $\hat{y}$ and all the points in the target point cloud $X$. In other words, if any of the $k$-NN of $\hat{y}$ to $X$ also have $\hat{y}$ in any of their $k$-NN, then the point $\hat{y}$ already has a \textit{good fit} and does not receive a beam-gap penalty.

\section{Implementation Details}
\subsection{Iterative Coarse-to-Fine Optimization}
The optimization process has two main coarse-to-fine aspects: (i) periodically up-sampling the mesh resolution (\emph{i.e.,} number of mesh elements); and (ii) increasing the number of point samples taken from the reconstructed mesh. 

Mesh subdivision, in the context of \ourmethod{}, has two main purposes: enabling the expression of richer and finer geometric details, and allowing the optimization to move away from local minima. Finer meshes are more flexible, since they can easily move toward their target. However, starting with a large mesh resolution will inevitably over-complicate the optimization process. To this end, the initial mesh starts out with a relatively small number of faces (roughly a couple thousand), and the optimization begins to move the mesh vertices to achieve a rough alignment of the starting mesh with the target point cloud. As the optimization progresses, the mesh is subdivided and smoothed, as it continues to deform into the target shape.

During the coarse optimization iterations, we upsample the mesh using the robust watertight manifold surface generation of Huang et al.~\shortcite{huang2018robust}, which we will refer to as $\mathsf{RWM}$. First, the network weights update and start to deform the initial coarse mesh to move toward the target. After $K$ optimization iterations (a hyperparameter, we use $K=1000$), the current deformed mesh is passed to $\mathsf{RWM}$. The new $\mathsf{RWM}$-generated mesh is a manifold, watertight and non-intersecting surface, which is used as the initial mesh to the next level of optimization. We use a high octree resolution (\emph{i.e.,} number of leaf nodes) in $\mathsf{RWM}$, to preserve the details recovered in the optimization and also introduce more polygons. The number of polygons in the $\mathsf{RWM}$-output will be simplified to the target number of faces that we pre-defined for the next optimization iteration. In practice, we increase the number of faces by $1.5$ after reaching the end of an optimization, and stop increasing the resolution once the maximum number of faces is met. Note that the weights and the constant random vector $C$ are re-initialized at the beginning of each optimization level.

The second aspect in the coarse-to-fine optimization is the number of points that are sampled on the reconstructed mesh. Initially, over-sampling the reconstructed mesh with too many points may significantly slow down the optimization process, since different points sampled from the same mesh face may (initially) be driven to vastly different directions by the loss function. 
To facilitate desirable convergence, we iteratively increase the number of reconstructed mesh point samples in each level of optimization.
Specifically, we pre-define: the number of starting samples $R_0$ (usually $R_0=15$k) and a final number of samples $R_K$. The number of reconstructed mesh samples increases linearly until reaching the maximum number $R_K$ after $K$ optimization iterations. After moving to the next optimization level, we restart the number of samples to $R_0$. For intricate shapes, we sample a maximum of ~$50$k-$100$k points, while simpler shapes require around $15$k-$30$k points.

The manifold surface generation helps the optimization in two ways. First, creating a clean non-intersecting surface makes it simpler for the optimization to displace the mesh vertex positions. Secondly, since it uses an octree to reconstruct the surface, the upsampled mesh has relatively uniformly sized triangles, which helps large flat areas that cover unentered cavities contain enough resolution to deform. The watertight manifold implementation is relatively fast, taking $10.5$ seconds on a single GTX 2080 TI GPU for a mesh with 40,000 triangles using 20,000 leaf nodes in the octree.

\subsection{Part-Based}
The GPU memory needed to optimize a given mesh increases as the mesh resolution increases. To alleviate this problem (at finer resolutions), a \partmesh{} data structure is defined, which enables the entire mesh to be optimized by parts. A \partmesh{} is a collection of sub-meshes which together make up the complete mesh. 

Given a reconstructed mesh $\hat{M}$, a \partmesh{} representing $\hat{M}$ will have a set of sub-meshes, where each contains a subset of the vertices and faces from $\hat{M}$. 
Once defined, the architecture of our network can be manipulated to receive a \partmesh{} as input, performing a forward pass on each part separately, accumulating the gradients for the entire mesh $\hat{M}$, and then performing back-propagation once using all the accumulated gradients. The final mesh reconstruction $\hat{M}$ can then be recovered from the \partmesh{}. 

Moreover, we enable overlapping regions between different sub-meshes in the \partmesh{}. In the case that a particular vertex is present in more than one sub-mesh, the final vertex position will be the average over all of the locations in each sub-mesh. The mesh is divided into sub-parts by splitting the vertices into an $n \times n$ grid.

\subsection{Initial Mesh Approximation} 
\label{sec:initalmesh}
Depending on the requirements of the reconstruction, we can use different methods for approximating the initial deformable mesh. For shapes known to have genus-0, we use the convex hull as the initial mesh. If the genus is unknown or greater than 0, we approximate the initial mesh through a series of steps. First, we calculate the the alpha shape or Poisson reconstruction from the input point cloud. Then, we apply the watertight manifold algorithm~\cite{huang2018robust} using a very coarse resolution octree (on the order of a few hundred leaf nodes). 

This coarse approximation has two main purposes: (i) it distances the initial mesh from potentially incorrect reconstructions, which are the result of noise, low-density regions, or incorrect normals (in the case of Poisson) and (ii) it helps close topologically incorrect holes that might occur in low-density regions (\emph{e.g.,} Figure~\ref{fig:genusn}).

\input{figures/06_experiments/069_genus/figure.tex}
\input{figures/noisy/harmonics}

\subsection{Run-time}
\input{figures/06_experiments/0610_real_scans/figure.tex}
The speed of our non-optimized  PyTorch~\cite{paszke2017automatic} implementation is a function of mesh resolution and number of samples. An iteration (forward and backward pass) takes $0.23$ seconds for a mesh with 2,000 faces and 15,000 points in the target and reconstructed mesh, or $0.59$ seconds for a mesh with  6,000 faces using a GTX 1080 TI graphics card. For a mesh with 40,000 faces (using $8$ parts), it takes $4.7$ seconds per iteration (\emph{i.e.,} $0.59$ seconds per part). The number of iterations required varies according to the complexity of the model, where simple models can converge in less than 1,000 iterations, while more complex ones may need around 10,000 (or more) iterations. Therefore, depending on the complexity, the total runtime for most models can range anywhere from a few ($\sim3$) minutes to several ($\sim3$) hours. While extremely high resolution results can take over 14 hours. A table with the runtimes for each of the models is listed in the supplementary material.

%% file: figures/06_experiments/068_self_similarity/anky.tex
\begin{figure*}[h]
    \centering
    \newcommand{\pl}{-3}
    \begin{overpic}[width=18cm]{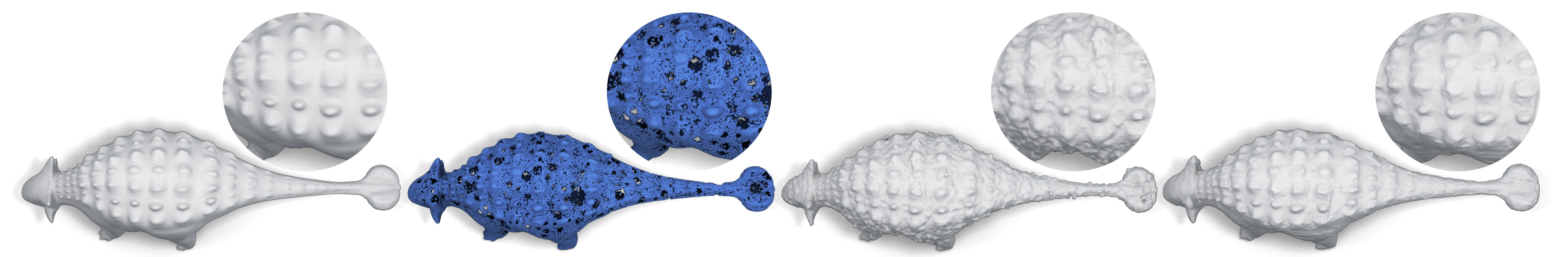}
    \put(6,  \pl){\textcolor{black}{Ground-Truth}}
    \put(33.7, \pl){\textcolor{black}{Input}}
    \put(55.2, \pl){\textcolor{black}{Smooth-prior}}
    \put(81, \pl){\textcolor{black}{Self-prior}}
    \end{overpic}
    \vspace{-5pt}
    \caption{The input point cloud is sampled from a (ground-truth) mesh, with added noise and missing regions. A smooth-prior reconstructs the surface locally, oblivious to the global shape. While the self-prior retains the reoccurring ridges in the back of the ankylosaurus and it smooths bumps which originated from noise. Note the smooth reconstruction along the tail and side of the body.}
    \label{fig:anky}
\end{figure*}

%% file: figures/06_experiments/068_self_similarity/flat_missing.tex
\begin{figure}[b!]
    \centering
    \newcommand{\pl}{-3}
    \begin{overpic}[width=\columnwidth]{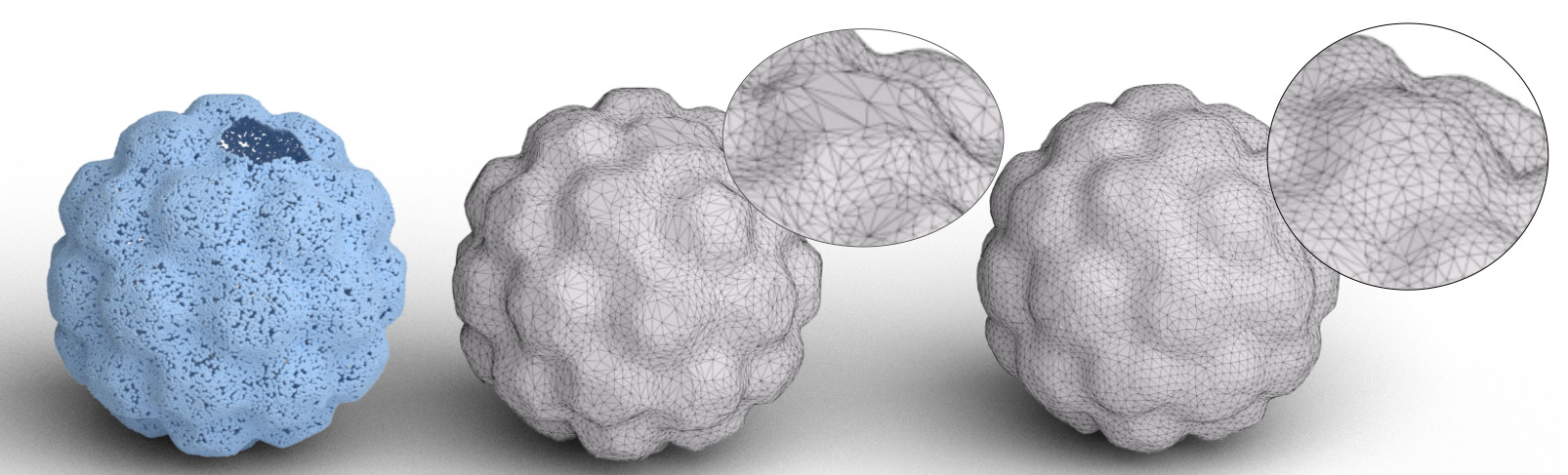}
    \put(12,  \pl){\textcolor{black}{Input}}
    \put(31, \pl){\textcolor{black}{Smooth-prior}}
    \put(68, \pl){\textcolor{black}{Self-prior}}
    \end{overpic}
    \vspace{-5pt}
    \caption{\ourmethod{} exploits self similarity in order to reconstruct a better missing half dome.}
    \label{fig:flat_missing}
\end{figure}

%% file: figures/06_experiments/062_converge/figure.tex
\begin{figure*}
\setlength\tabcolsep{0pt} % default value: 6pt
\newcommand{\wf}{16}
\newcommand{\pl}{-1.5}
\includegraphics[width=\wf cm]{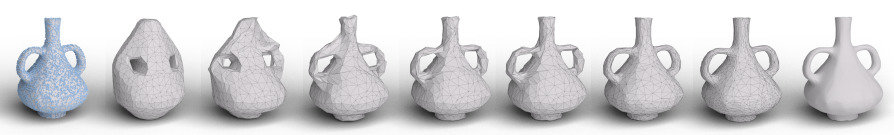} \\
\includegraphics[width=\wf cm]{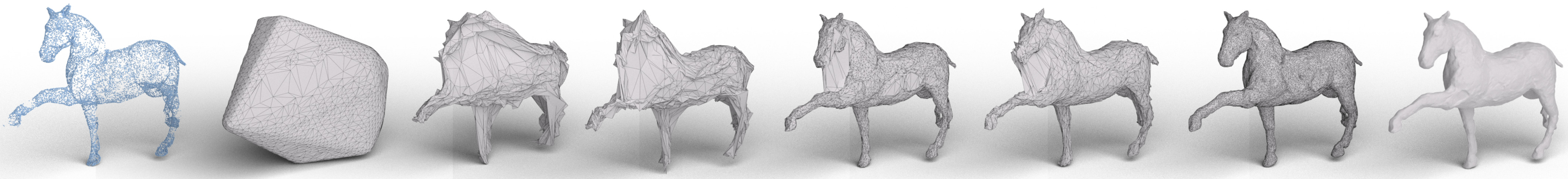} \\
\includegraphics[width=\wf cm]{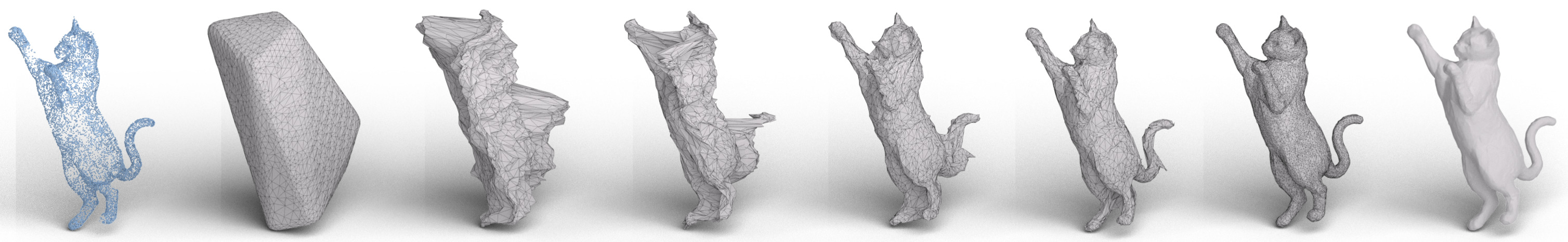} \\ \includegraphics[width=\wf cm]{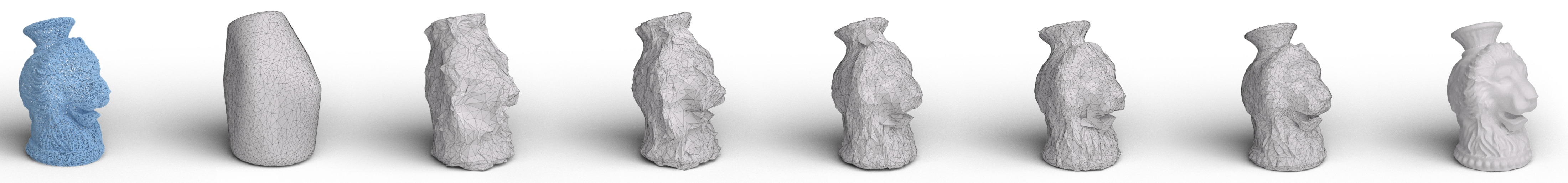} \\
\includegraphics[width=\wf cm]{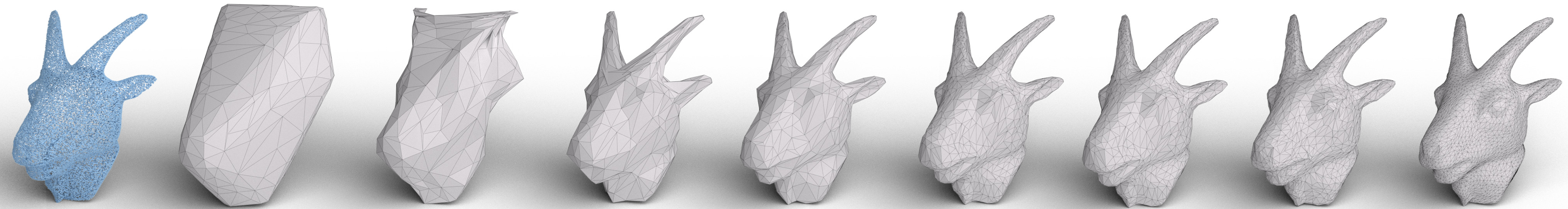} \\
\caption{Iterative progression through the \ourmethod{} optimization process. Starting with the input point cloud (left), the inital mesh begins to deform towards the input point cloud. The mesh is iteratively upsampled, to add finer details throughout the optimization procedure.}
\label{fig:convergence}
\end{figure*}

%% file: figures/06_experiments/069_genus/figure.tex
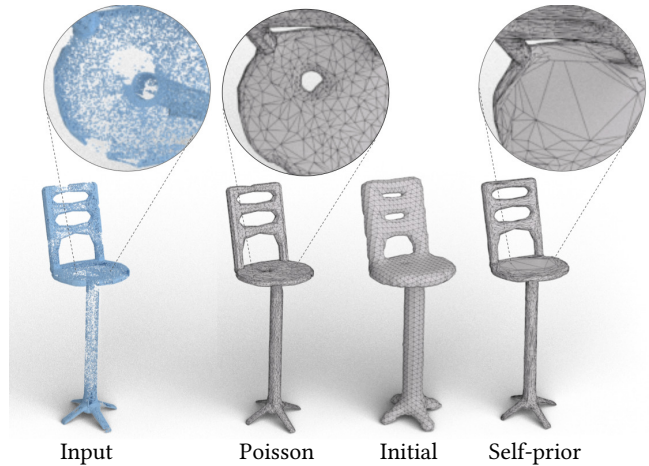
\begin{figure}
    \centering
    \newcommand{\wf}{8.5}
    \newcommand{\pl}{-3}
    \begin{overpic}[width=\wf cm]{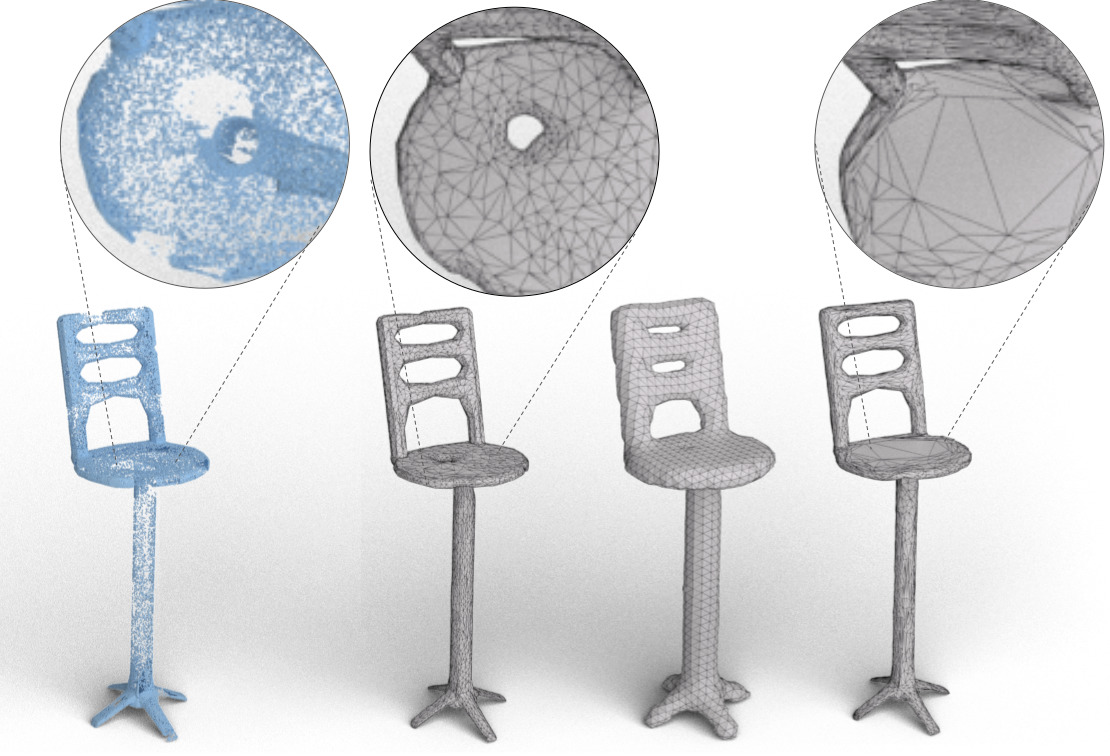}
    \put(8,  \pl){\textcolor{black}{Input}}
    \put(36,  \pl){\textcolor{black}{Poisson}}
    \put(58, \pl){\textcolor{black}{Initial}}
    \put(75, \pl){\textcolor{black}{Self-prior}}
    \end{overpic}
    \caption{Arbitrary genus with missing portions. Starting with the input point cloud, Poisson reconstruction results in incorrect topological holes. We apply a low resolution octree tree to close incorrect holes and create a coarse initial mesh which is used as the initial mesh. The result of \ourmethod{} is a topologically correct mesh which preserves the details from the input point cloud, and does not overfit to incorrect low density regions (note the top of the chair).}
    \vspace{-5pt}
    \label{fig:genusn}
\end{figure}

%% file: figures/noisy/harmonics.tex
\begin{figure}[b]
    \centering
    \newcommand{\pl}{-3}
    \begin{overpic}[width=\columnwidth]{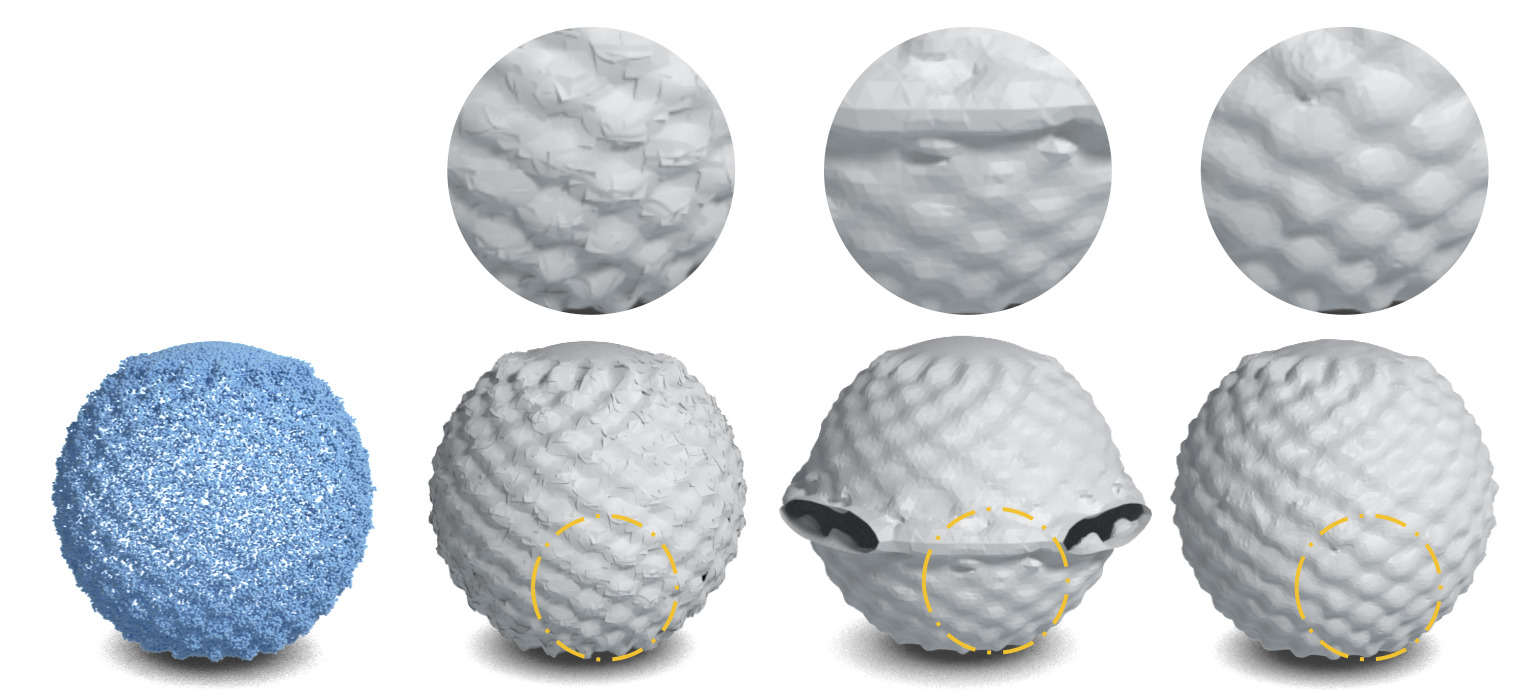}
    \put(10, \pl){\textcolor{black}{Input}}
    \put(35, \pl){\textcolor{black}{DGP}}
    \put(52, \pl){\textcolor{black}{DGP+Poisson}}
    \put(81, \pl){\textcolor{black}{Self-Prior}}
    \end{overpic}
    \vspace{-5pt}
    \caption{Surface reconstruction on noisy input with self-repetitions. Sampling the local DGP charts which are passed as input to Poisson (which struggles with unoriented normals).}
    \label{fig:dgpcomp}
\end{figure}
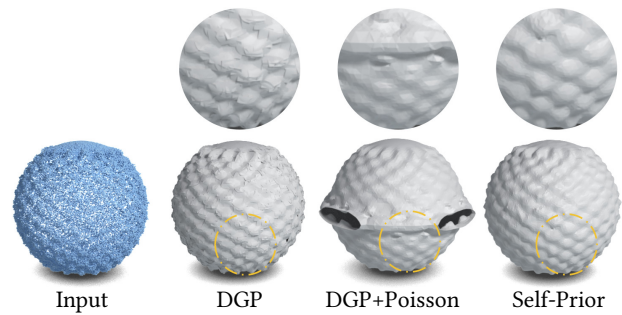

%% file: figures/06_experiments/0610_real_scans/figure.tex
\begin{figure*}[h]
\setlength\tabcolsep{0pt} % default value: 6pt
\newcommand{\wf}{18}
\includegraphics[width=12 cm]{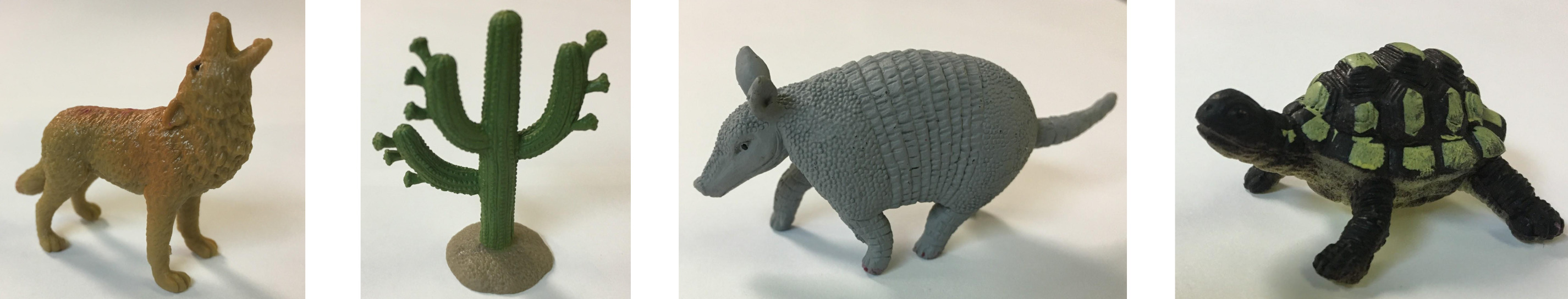} \\
\includegraphics[width=\wf cm]{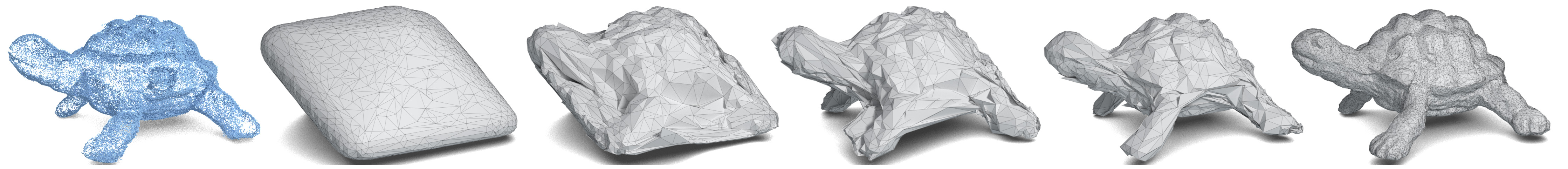}\\
\includegraphics[width=\wf cm]{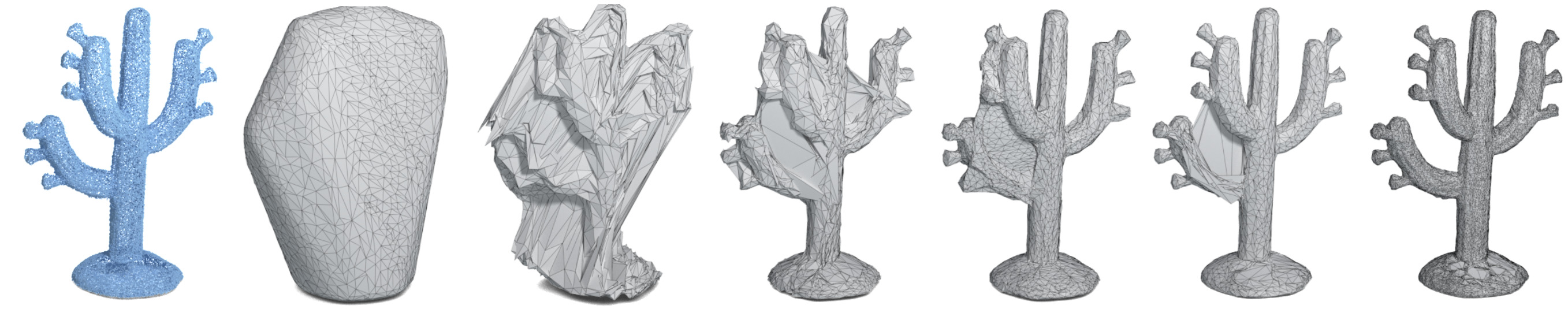}\\
\includegraphics[width=\wf cm]{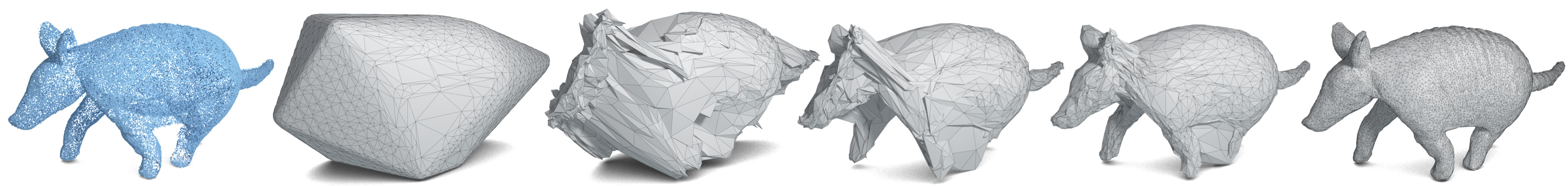} \\
\includegraphics[width=\wf cm]{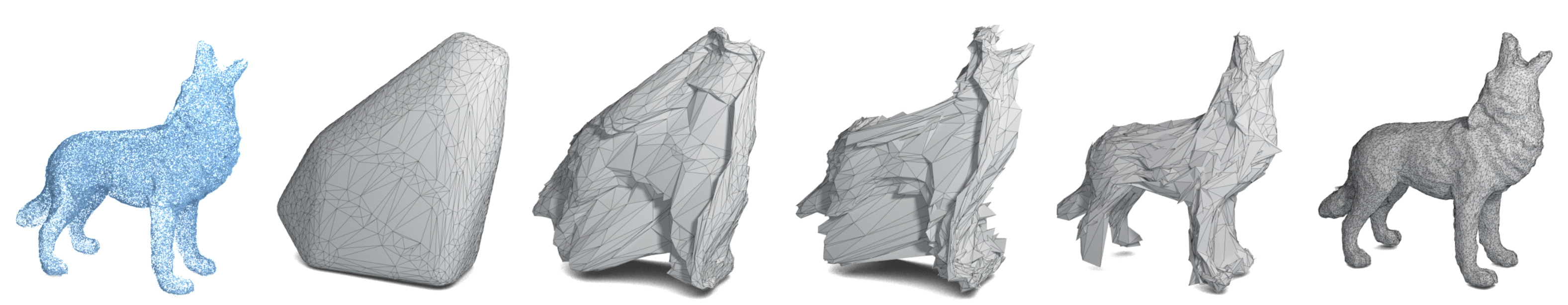} \\
\caption{Mesh reconstruction process of four objects (top row) scanned using NextEngine 3D laser scanner.}
\label{fig:scans}
\end{figure*}

%% file: experiments.tex
\section{Experiments}
We validate the applicability of \ourmethod{} through a series of qualitative and quantitative experiments involving shapes with missing regions, noise and challenging cavities. 
We start with qualitative results of our method on real scanned objects from~\cite{Choi2016} as well as from the NextEngine 3D laser scanner. We provide additional results and quantitative experiments on point sets sampled from a ground-truth mesh surface. These mesh datasets include: Thingi10k~\cite{Thingi10K}, COSEG~\cite{wang2012active}, TOSCA high-resolution~\cite{bronstein2006efficient} and the Surface Reconstruction Benchmark~\cite{berger2013benchmark}. 

\input{figures/06_experiments/065_unoriented/figure.tex}
\input{figures/06_experiments/064_choi_scan/figure.tex}
\subsection{Real Scans}

Before presenting our results on real scans, we discuss the problem of imperfect normals, which is a characteristic of real scanned data.
While our method uses the normal information provided in the point cloud, as we demonstrate below, it is robust to noisy and inaccurate normals. This is due to the small weight on the normal penalty (cosine similarity) and the self-prior.

% {\bf Normal Estimation.}
\subsubsection*{Normal Estimation}
Generally, in ideal conditions, points uniformly sample the shape, are noise-free, have correct / oriented normals and have sufficient resolution. Under such ideal conditions, a smooth-prior (\emph{e.g.,} Poisson reconstruction) is an excellent choice for reconstruction. An advantage of \ourmethod{} compared to smooth-priors is the ability to cope with unoriented normals. Meaning the \emph{line} defining the normal is correct, however,  the direction (inward or outward) may be be incorrect. 
If normals are unoriented (or overly noisy), it is a rather challenging task to correctly orient them with existing tools. 

To evaluate the orientation problem, we sample a mesh without normals, and estimate the normals and their orientation using~\cite{Zhou2018}. The results of our approach are presented in Figure~\ref{fig:unoriented}. The input point cloud is visualized with a heatmap of the angle of the estimated normal error, and the results of Screened Poisson reconstruction and \ourmethod{} are displayed side-by-side. Note that the loss function that we use aims to align the normals of the input point cloud and its corresponding point in the mesh using a cosine similarity. In the event that the normals are unoriented, we use the absolute dot-product between the two vectors, which gives equal penalty for vectors $180^{\circ}$ out of phase.  

% {\bf NextEngine 3D scans.} 
\subsubsection*{NextEngine 3D scans}
We have scanned four objects using the NextEngine 3D laser scanner and show the reconstruction results and iterative convergence in Figure~\ref{fig:scans}. 
The second column to the left shows the initial mesh, which is the convex hull.
Observe that despite the large holes in the input point cloud, \ourmethod{} results in a plausible reconstruction.

% {\bf \citet{Choi2016} scans.}
\subsubsection*{\citet{Choi2016} scans}
Figure~\ref{fig:noisy} demonstrates the reconstruction from real scans obtained from~\cite{Choi2016}. Not all normals in these scans are oriented to the correct direction, and the smooth reconstruction therefore has some observable artifacts which are not present in \ourmethod{} results. 

\subsection{Accuracy Metric}
For a quantitative comparison, we evaluate \ourmethod{} on a point cloud obtained from a ground-truth mesh. For the metric between the ground-truth and the reconstructed mesh, we use the F-score metric as discussed in Knapitsch et al.~\shortcite{Knapitsch2017}. 
The F-score is defined by the precision $P(\tau)$ and recall $R(\tau)$ at a given distance threshold $\tau$. 
The precision can be thought of as some type of accuracy of the reconstruction, which considers the distance from the reconstructed mesh to ground-truth mesh. The recall provides an indication of how well the reconstructed mesh \emph{covers} the target shape, which considers the distance from the ground-truth mesh to the reconstructed mesh. The F-score $F(\tau)$ is the harmonic mean between $P(\tau)$ and $R(\tau)$. The best F-score is $100$ and the worst is $0$. For more details about the F-score, refer to the supplementary material.
It is worth noting that like all \emph{distortion}-based metrics for inverse problems, a high F-score may not necessarily imply better visual quality, and vice-versa~\cite{blau2018perception}.

\input{figures/06_experiments/060_noise_benchmark/060_noise_benchmark.tex}
\input{figures/06_experiments/060_noise_benchmark/060_noise_benchmark_qual.tex}

\subsection{Comparisons}
% {\bf Denoising.} 
\subsubsection*{Denoising}
To evaluate our approach on the task of denoising, we use two meshes as ground-truth for performing quantitative comparisons. We sample each mesh with $75,000$ points and add a small amount of Gaussian noise to each $<x, y, z>$ coordinate in the input point cloud. We sample from the noise distribution five times per shape and report F-score statistics for the five versions of each shape in~Table~\ref{table:noise_benchmark}. Qualitative examples are shown in Figure~\ref{fig:noise_benchmark}.
% \rev{(where $\tau$ is for $Q=100$, see supplementary for details on Q)}

We compare against Screen Poisson surface reconstruction~\shortcite{kazhdan2013screened}, DGP~\shortcite{williams2019deep} and PointCleanNet (PCN)~\shortcite{rakotosaona2019pointcleannet}, which trains a neural network directly for the task of point cloud denoising. Note that we apply a Poisson reconstruction on the result of PCN for the sake of visualizing their result, but the F-score is computed on the raw (cleaned) point samples. We use the same configuration for each method across all shapes.
Note that the deep neural network based approaches outperform the classic Poisson smooth-prior. A qualitative comparison on noisy data for a shape with self-repetition is shown in Figure~\ref{fig:dgpcomp}. The local charts generated by DGP are sampled and used as input to Poisson.
\input{figures/06_experiments/063_inpaintint_benchmark/figure.tex}

% {\bf Low Density Completion.} 
\subsubsection*{Low Density Completion}
To evaluate our approach on the task of \emph{shape completion}, we use two meshes as ground-truth for performing quantitative comparisons. In this task, we aim to complete the missing parts of a shape, which contains large regions with very little to no samples. In order to acquire such data, the input point cloud was sampled from meshes where several large random regions were removed. We define a probability for: the number and size of the low-density regions and the frequency for sampling inside them. 

We sample the missing region probabilities five times per shape and report statistics for the five versions of each shape in Table~\ref{table:completion_benchmark}. Qualitative examples are shown in Figure~\ref{fig:completion_benchmark}. 
In the case of \emph{completion}, we modify the \textit{ground-truth} used in the F-score computation to better reflect how well the missing regions were completed.
The precision component remains with respect to the entire ground-truth surface, while the recall is with respect to the (ground-truth) missing portion of the surface only. Since recall gives some notion of coverage, this indicates how well the \emph{missing} regions were \textit{covered}. Note that since we used the convex-hull as the initial mesh, this implicitly provides the correct genus to our approach. In this sense, the comparison is not entirely fair, since the other approaches cannot use this information which provides an advantage for hole filling.
\input{figures/06_experiments/063_inpaintint_benchmark/quant.tex}
 
\subsection{Network is a Prior}
\input{figures/06_experiments/061_G_convergence/061_G_convergence.tex}
In order to separate the effect of the proposed reconstruction criteria and coarse-to-fine framework from the effect of the self-prior (\emph{i.e.,} network weights), we run an ablation to demonstrate the importance of the network. We optimize the same criteria directly through back-propagation in the same coarse-to-fine fashion (without a network). We call this setting \emph{direct} optimization, since it directly computes the ideal deformable mesh vertex placements given the same criteria, through back-propagation. 

Figure~\ref{fig:g} shows that the self-prior (provided by the network) plays a crucial role in facilitating a desirable convergence:
The proposed objective for reconstruction (\emph{i.e,} Chamfer and beam-gap), in addition to our coarse to fine framework (\emph{i.e.,} iterative subdivision) are not sufficient alone to generate the desirable results.
This indicates that while a direct optimization approach is prone to local minima, the network as a prior can avoid these minimas by its weight sharing and self similarity capabilities.

\begin{figure}
    \newcommand{\pl}{-3}
    \begin{overpic}[width=8.5cm]{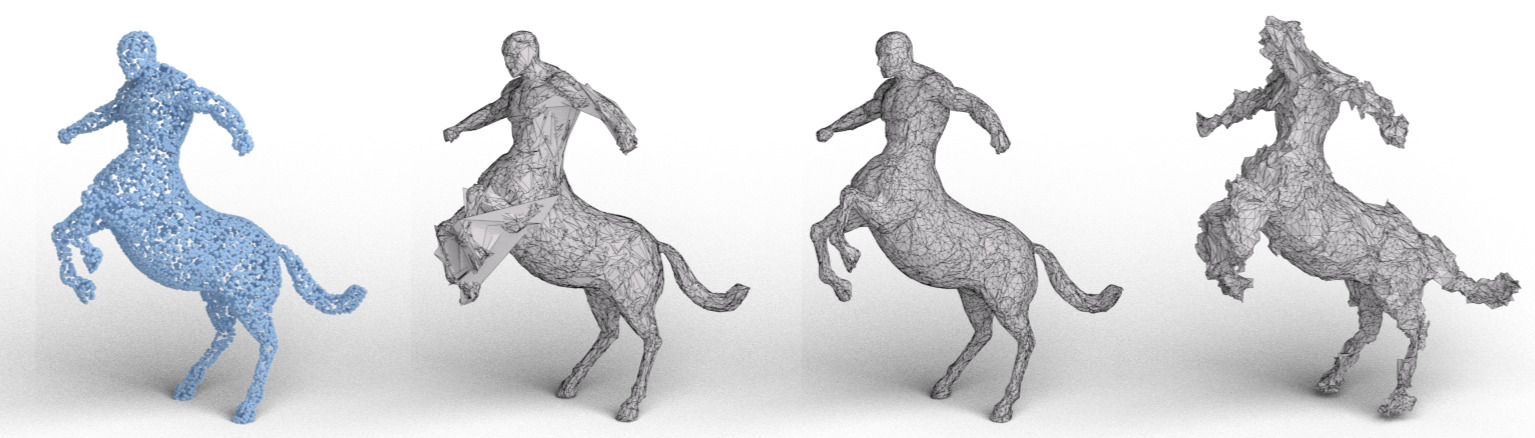}
    \put(6,  \pl){\textcolor{black}{Input}}
    \put(26,  \pl){\textcolor{black}{Optimization}}
    \put(51, \pl){\textcolor{black}{Over-param}}
    \put(76, \pl){\textcolor{black}{Under-param}}
    \end{overpic}
    \caption{Representational power of network priors. Deforming a mesh to the input point cloud using direct optimization (no network prior) leads to undesirable results. \ourmethod{} relies on the representation power of deep CNNs to deform the input mesh, \emph{i.e.,} an overparameterized (\emph{Over-param}) network, where as an \textit{underparameterized} (\emph{Under-param}) network is not expressive enough to produce favorable results.}
    \label{fig:network_rep_prior}
\end{figure}

Our network architecture is a U-Net configuration with residual and skip connections. The network hyper-parameters define the representational power of the self-prior. This translates to the effectiveness of the self-prior, since a \emph{powerful} network architecture should lead to a strong self-prior. While the notion of what makes a neural network effective is somewhat obscure, empirically neural networks tend to work poorly in under-parameterized settings (\emph{i.e.,} with not enough trainable weights). To test this, we ran $50$ configurations of the self-prior hyperparameters and manually classified the results as either a \textit{noisy} or \textit{detailed} on the centaur shape (Figure~\ref{fig:network_rep_prior}). We present the average of various parameters from each of the classes in the table below. The clean class was generated from 
\begin{wrapfigure}{l}{0.2\textwidth}
\begin{tabular}{l c c}
\hline
 & noisy & detailed \\ [0.5ex] % inserts table
%heading
\hline % inserts single horizontal line
F-score & 81.24\% & 99.9\% \\
feat & 11.6 & 97.8 \\
minfeat & 4 & 21.6 \\
params & 6.5k & 263k \\
pool & 40\% & 20\% \\
res & 1.9 & 2 \\
\hline
\end{tabular}
\label{table:abalation}
\end{wrapfigure}
configurations with more deep features in the bottleneck of the U-Net architecture (\emph{feat}) and a larger number of learned parameters (\emph{params}). Narrowing the convolutional width (\emph{minfeat}) results in noisy reconstructions.
Moreover, pooling (\emph{pool}) less features also seemed to improve performance, while the number of residual connections (\emph{res}) did not seem to have an effect. Note that we observed empirically, specifically for the task of shape completion, that increased pooling had a more desirable outcome with respect to completion.

%% file: figures/06_experiments/065_unoriented/figure.tex
\begin{figure*}
    \centering
    \newcommand{\pl}{-2}
    \begin{overpic}[width=15cm]{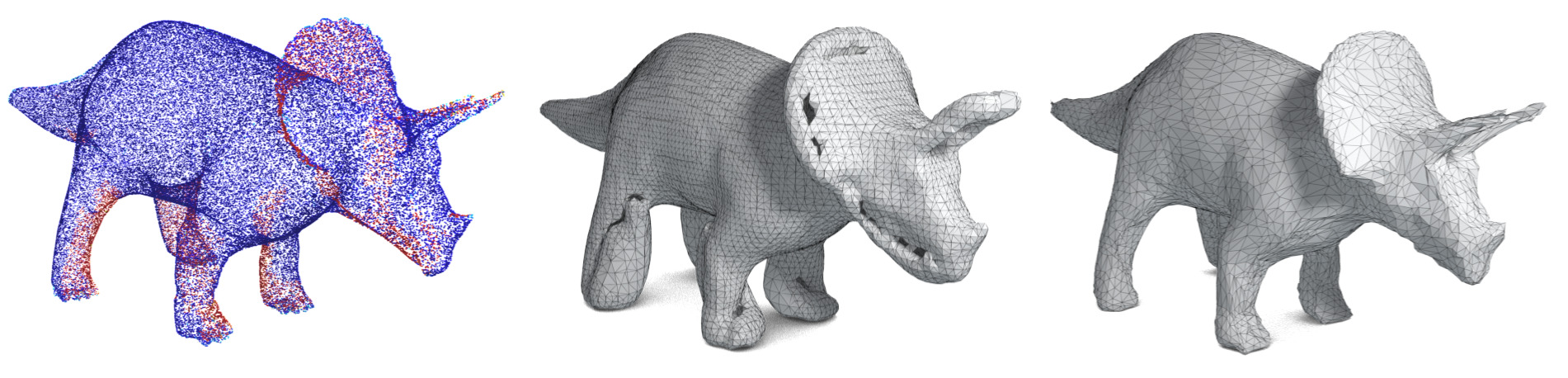}
    \put(12,  \pl){\textcolor{black}{Input}}
    \put(44, \pl){\textcolor{black}{Poisson}}
    \put(78, \pl){\textcolor{black}{Ours}}
    \end{overpic}
    \caption{Reconstruction results on estimated normals. Calculating normals on the input point cloud and applying a normal orientation algorithm (heat map colors of error angle to ground-truth normal). While Screened Poisson is sensitive to unoriented normals, \ourmethod{} is agnostic to orientation.}
    \label{fig:unoriented}
\end{figure*}

%% file: figures/06_experiments/064_choi_scan/figure.tex
\begin{figure}[b]
    \centering
    \newcommand{\pl}{-3}
    \begin{overpic}[width=8.5cm]{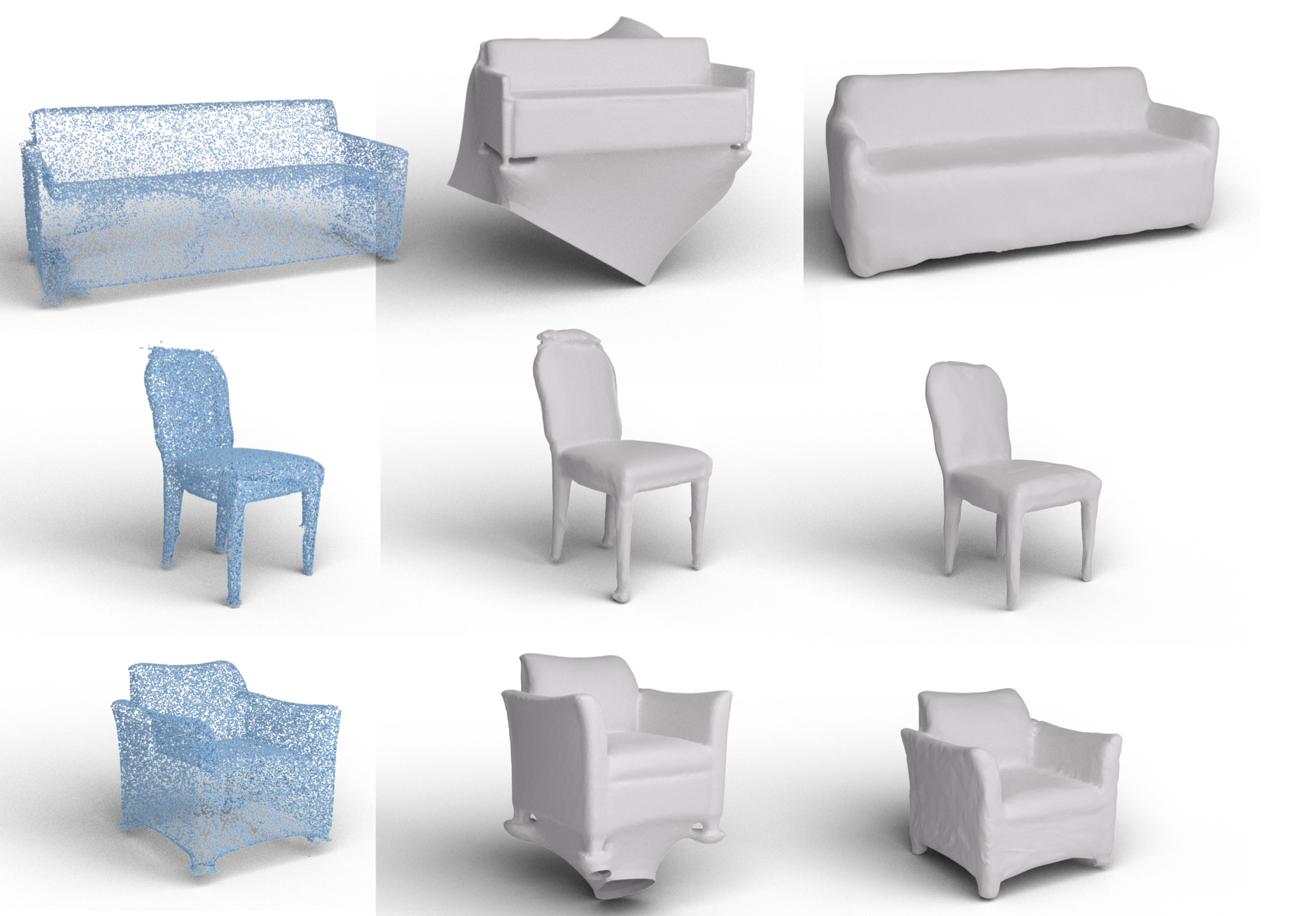}
    \put(12,  \pl){\textcolor{black}{Input}}
    \put(41, \pl){\textcolor{black}{Poisson}}
    \put(72, \pl){\textcolor{black}{Self-prior}}
    \end{overpic}
    \caption{Reconstruction from real scans from~\cite{Choi2016}. Note that \ourmethod{} guarantees reconstruction of a watertight surface despite noise, missing regions and unoriented normals.}
    \label{fig:noisy}
\end{figure}

%% file: figures/06_experiments/060_noise_benchmark/060_noise_benchmark.tex
\begin{table}[b]
\caption{Shape denoising comparison. F-score (larger is better) statistics for five different noise samples of each underlying shape.}
\vspace{-5pt}
\begin{tabular}{l c c c c | c c c c}
\hline
& \multicolumn{4}{c}{\textbf{Guitar}} & \multicolumn{4}{c}{\textbf{Tiki}} \\
\hline\hline
\textbf{Method} & avg & std & min & max & avg & std & min & max \\ [0.5ex]
\hline
Poisson
& 86.4 & 2.6 & 82.5 & 89.1
& 47.6 & 0.3 & 47.2 & 47.9 \\
PCN 
& 97.2 & 1.0 & 95.8 & 98.3
& 58.6 & 0.4 & 57.8 & 59.1 \\
DGP 
& 92.9 & 1.3 & 90.9 & 94.2
& 50.7 & 0.7 & 49.8 & 51.9 \\
Ours
& 98.3 & 0.5 & 97.5 & 98.8
& 60.2 & 0.4 & 59.6 & 60.6 \\ [1ex]
\hline
\end{tabular}
\label{table:noise_benchmark}
\end{table}

%% file: figures/06_experiments/060_noise_benchmark/060_noise_benchmark_qual.tex
\begin{figure}[]
\setlength\tabcolsep{0pt} % default value: 6pt
\newcommand{\wf}{8.5}
\newcommand{\pl}{-3}
\includegraphics[width=\wf cm]{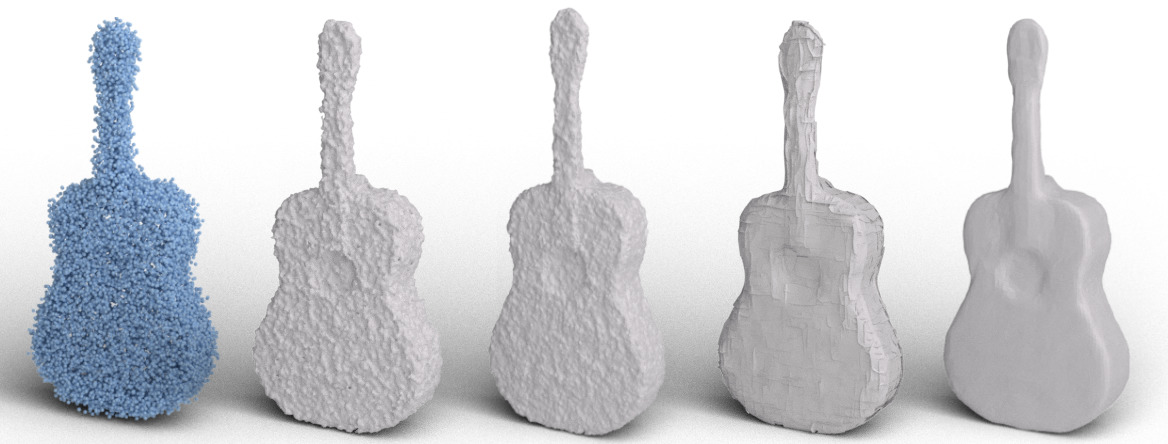} \\
\begin{overpic}[width=\wf cm]{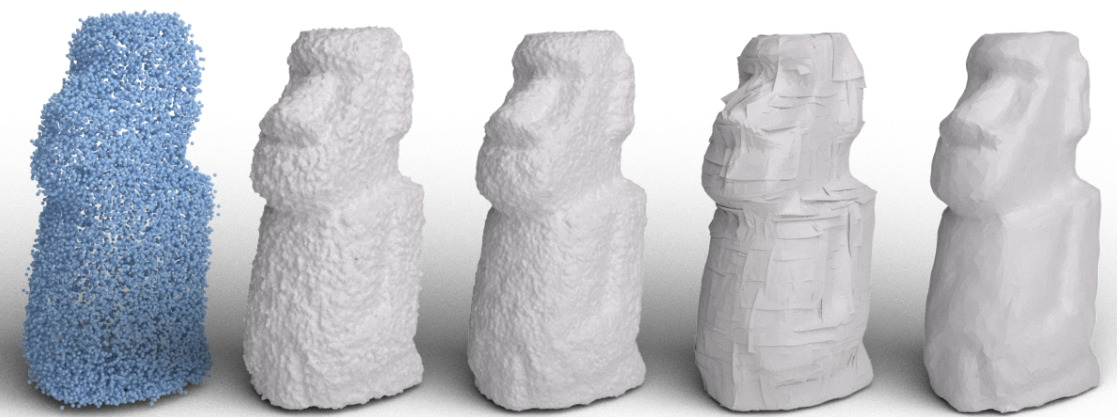}
    \put(8,  \pl){\textcolor{black}{Input}}
    \put(25, \pl){\textcolor{black}{Poisson}}
    \put(45, \pl){\textcolor{black}{PCN}}
    \put(68, \pl){\textcolor{black}{DGP}}
    \put(86, \pl){\textcolor{black}{Ours}}
\end{overpic}
\caption{Qualitative results from the noisy comparison.}
\label{fig:noise_benchmark}
\end{figure}

%% file: figures/06_experiments/063_inpaintint_benchmark/figure.tex
\begin{figure}[b]
\setlength\tabcolsep{0pt} % default value: 6pt
\newcommand{\wf}{8.5}
\newcommand{\pl}{-2.5}
\includegraphics[width=\wf cm]{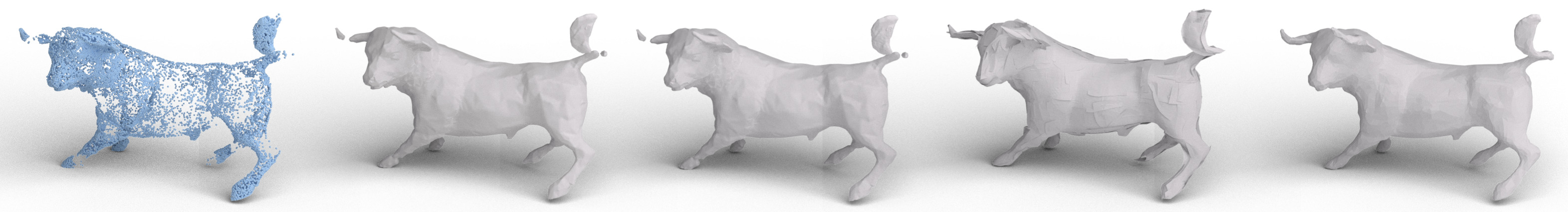} \\
\begin{overpic}[width=\wf cm]{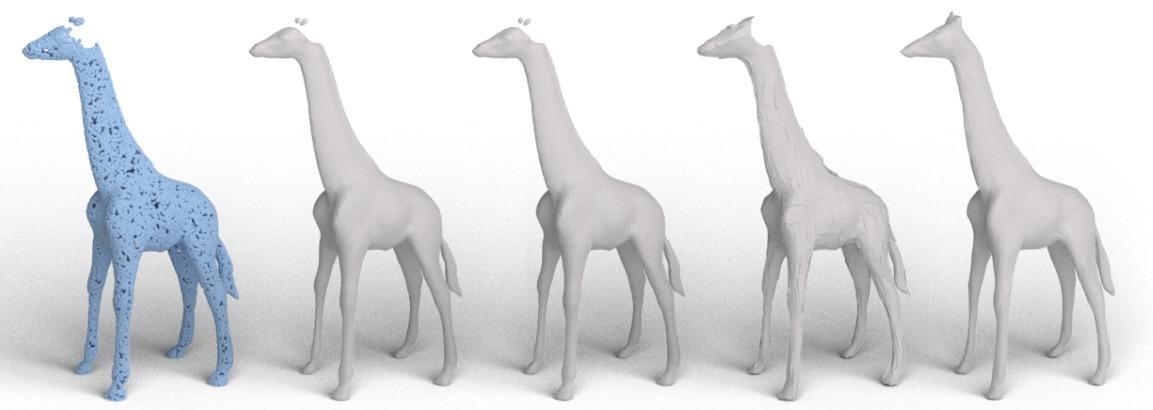}
    \put(10,  \pl){\textcolor{black}{Input}}
    \put(29, \pl){\textcolor{black}{Poisson}}
    \put(50, \pl){\textcolor{black}{PCN}}
    \put(69, \pl){\textcolor{black}{DGP}}
    \put(87, \pl){\textcolor{black}{Ours}}
\end{overpic}
\caption{Qualitative results on shape completion comparison.}
\label{fig:completion_benchmark}
\end{figure}

%% file: figures/06_experiments/063_inpaintint_benchmark/quant.tex
\begin{table}
\caption{Shape completion comparison. F-score (higher is better) statistics for different missing portions of the underlying shape ($\times 5$ per shape). The precision is reported for the entire shape, while the recall is with respect to the missing portion only.}
\vspace{-5pt}
\begin{tabular}{l c c c c | c c c c}
\hline
& \multicolumn{4}{c}{\textbf{Giraffe}} & \multicolumn{4}{c}{\textbf{Bull}} \\
\hline\hline
\textbf{Method} & avg & std & min & max & avg & std & min & max \\ [0.5ex]
\hline
Poisson
& 58.3 & 12.4 & 41.2 & 74.4 
& 84.9 & 1.6 & 82.8 & 87.4 \\
PCN 
& 58.3 & 12.5 & 41.3 & 74.4 
& 84.5 & 1.7 & 82.3 & 87.2 \\
DGP
& 63.3 & 13.6 & 42.6 & 76.4 &
75.7 & 2.9 & 71.6 & 80.3 \\
Ours
& 80.7 & 11.5 & 58.2 & 90.8 
& 87.4 & 2.8 & 83.5 & 92.0 \\ [1ex]
\hline
\end{tabular}
\label{table:completion_benchmark}
\end{table}

%% file: figures/06_experiments/061_G_convergence/061_G_convergence.tex
\begin{figure}[b]
    \centering
    \begin{overpic}[width=8.5cm]{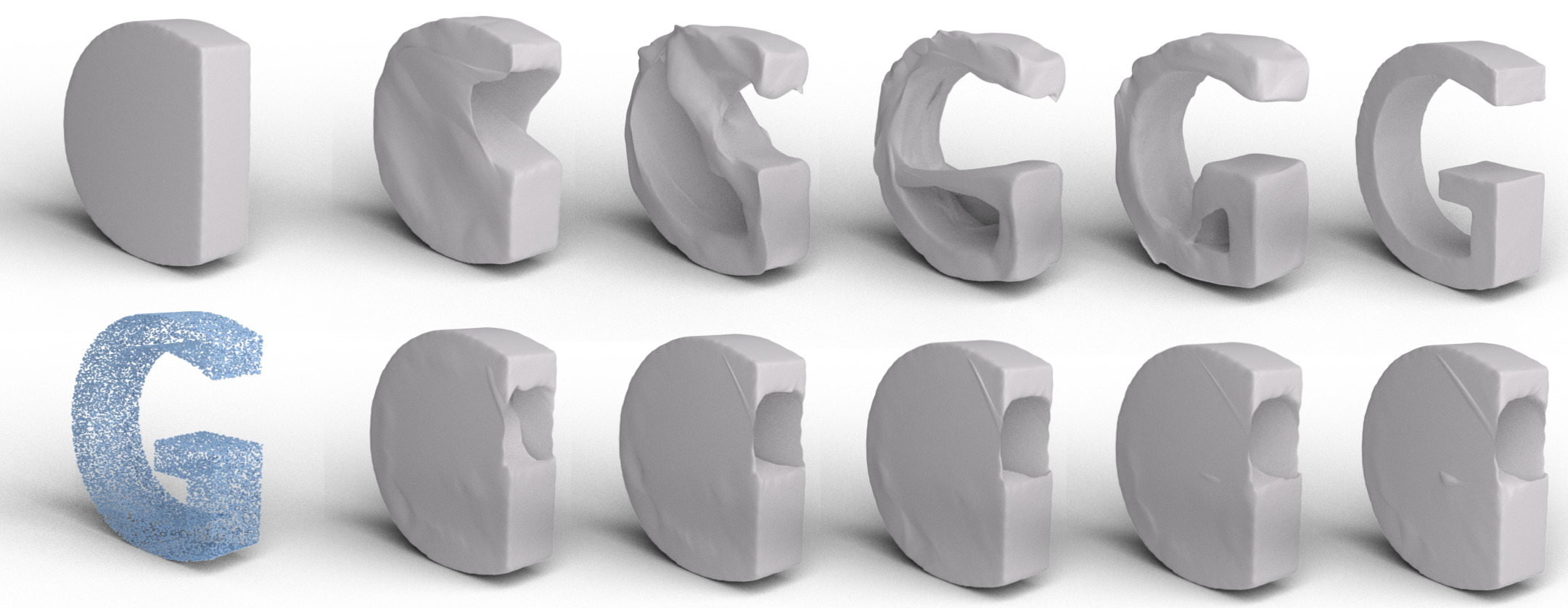}
    \put(2,22){\textcolor{black}{(a)}}
    \put(2,5){\textcolor{black}{(b)}}
    \put(20,22){\textcolor{black}{(c)}}
    \put(20,5){\textcolor{black}{(d)}}
    \end{overpic}
    \vspace{-10pt}
    \caption{Network is a prior. Starting with the initial mesh (a) and the target point cloud (b), the iterative \ourmethod{} convergence is shown along the upper row (c). Optimizing the same objective directly without a network gets trapped in a local minimum (d). }
    \label{fig:g}
\end{figure}

%% file: conclusion.tex
\section{Discussion}
The rapid emergence of neural networks has had a tremendous impact in the development of graphical processing, in particular for images, videos, volumes and other regular representations. The use of neural networks on irregular structures is still an open research problem. Most of the research has focused on point clouds, with only a few operating directly on irregular meshes, which focused on analysis only.
In this work, we developed a neural network to perform a regression task, which optimizes the geometry of an irregular mesh. The network learns to displace the vertices of the given mesh in the context of surface mesh reconstruction. Moreover, the strategy of deforming a given mesh preserves the genus and provides control, which is lacking in competing methods.

The key point of our presented method is that reconstruction is based on a self-prior that is learned during the mesh optimization itself, and enjoys the innate properties of the network structure, which we refer to as the self-prior. Central to the self-prior is the weight-sharing structure of a CNN, which inherently models recurring and correlated structures and, hence, is weak in modeling noise and outliers, which have non-recurring geometries. Fortunately, natural shapes have strong self-correlation across multiple scales. Their surface is typically piecewise smooth or may contain geometric textures which consist of recurring elements (\textit{e.g.}, most notable in Figures \ref{fig:teaser}, \ref{fig:selfsim}, \ref{fig:anky}, \ref{fig:flat_missing}). Thus, this self-prior excels in learning and modeling natural shapes, and,
in a sense, is \emph{magical} in completing missing parts and removing outliers or noise.

We demonstrate the applicability of our reconstruction method on imperfect point clouds that are noisy and have missing regions, and show that it performs well in the tasks of denoising and completion.  Despite these promising findings, we note the current limitations of our approach: akin to most optimization techniques, our method is quite expensive in terms of time and memory, compared to state-of-the-art surface reconstruction techniques. 
Furthermore, although we demonstrate handling shapes with genus greater than zero, we had to rely on available techniques which may produce an initial mesh with imperfect results, especially in scenarios with non-ideal conditions.

An interesting direction for future work, could involve developing a separate network that can detect the genus of point cloud without relying on training data: following our approach to learn a self-prior while processing the input data. Another interesting avenue coulbe involve exploring the possibility of using our approach to obtain compatible triangulations between shapes, for example, by optimizing the same deformable mesh to shrink-wrap two or more different objects.

\begin{acks}
We thank Daniele Panozzo for his helpful suggestions. We are also thankful for help from Shihao Wu, Francis Williams, Teseo Schneider, Noa Fish and Yifan Wang. We are grateful for the 3D scans provided by Tom Pierce and Pierce Design. This work is supported by the NSF-BSF grant (No. 2017729), the European research council (ERC-StG 757497 PI Giryes), ISF grant 2366/16, and the Israel Science Foundation ISF-NSFC joint program grant number 2472/17.
\end{acks}